\documentclass[english,pre]{achemso}
\usepackage[T1]{fontenc}
\usepackage[latin9]{inputenc}
\usepackage{amsbsy}
\usepackage{microtype}
\usepackage{amstext}
\usepackage{graphicx}
\usepackage{setspace}
\usepackage{esint}
 
\makeatletter
\usepackage{setspace}
\usepackage{lipsum}

\title{Droplet transport in a nanochannel coated by hydrophobic semiflexible
polymer brushes: the effect of chain stiffness}
\author{K. Speyer$^{\dag}$ and C. Pastorino}
\affiliation{$^{*}$Departamento de Física de la Materia Condensada, Centro Atómico
Constituyentes, CNEA, Av.Gral.~Paz 1499, 1650 Pcia.~de Buenos Aires,
Argentina}
\email{pastor@cnea.gov.ar}
\affiliation{$^{\dag}$ CONICET, Avenida Rivadavia 1917, C1033AAJ Buenos Aires,
Argentina}

\AbstractOn
\SectionNumbersOn 
\SectionsOn 

\makeatother

\usepackage{babel}
\begin{document}
\begin{abstract}
We study the influence of chain stiffness on droplet flow in a nano-channel,
coated with semiflexible hydrophobic polymers by means of non-equilibrium
molecular-dynamics simulations. The studied system is then a moving
droplet in the slit channel, coexisting with its vapor and subjected
to periodic boundary conditions in the flow direction. The polymer
chains, grafted by the terminal bead to the confining walls, are described
by a coarse-grained model that accounts for chain connectivity, excluded
volume interactions and local chain stiffness. The rheological, frictional
and dynamical properties of the brush are explored over a wide range
of persistence lengths. We find a rich behavior of polymer conformations
and concomitant changes in the friction properties over the wide range
of studied polymer stiffnesses. A rapid decrease in the droplet velocity
was observed as the rigidity of the chains is increased for polymers
whose persistence length is smaller than their contour length. We
find a strong relation between the internal dynamics of the brush
and the droplet transport properties, which could be used to tailor
flow properties by surface functionalization. The monomers of the
brush layer, under the droplet, present a collective ``treadmill
belt'' like dynamics which can only be present due the the existence
of grafted chains. We describe its changes in spatial extension upon
variations of polymer stiffness, with bidimensional velocity and density
profiles. The deformation of the polymer brushes due to the presence
of the droplet is analyzed in detail. Lastly, the droplet-gas interaction
is studied by varying the liquid to gas ratio, observing a 16\% speed
increase for droplets that flow close to each other, compared to a
train of droplets that present a large gap between consecutive droplets. 
\end{abstract}

\twocolumn
\linespread{1}
\singlespacing
\section{Introduction}

Different forms of liquid flow under high confinement regime has gained
attention in the last years due to the importance in a broad range
of areas, like biology\cite{doi:10.1146/annurev.bioeng.9.060906.151959,Lanotte_2014,CruzChu2014232},
medicine\cite{http://dx.doi.org/10.1038/nature13118} and biotechnology\cite{BIOT:BIOT201500278}.
Microfluidics deals with flows in the pico-liter scale, focusing on
accurate manipulation of liquids and gases in micro and nano-channels.
Lab-on-chip devices, where complex biochemical reactions are automated,
are an example of promising applications in this field. Although large-scale
industrial implementations are not available yet\cite{Volpatti2014347},
microfluidic devices are widely used in research laboratories \cite{12ref19H,BIOT:BIOT201500278,http://dx.doi.org/10.1038/nature13118}. 

A low Reynolds number ($Re$) and a high surface-area-to-volume ratio
are characteristics of micro and nano-fluidics. Under these conditions,
the flow is largely influenced by the confining substrate, which determines
the boundary conditions. The no-slip boundary condition, usually assumed
in macro scale calculations, is an approximation that is no longer
valid in the micro-scale\cite{Lauga2007}. Efforts are being made
to understand and control liquids in the picoliter scale, by changing
the properties of the inner coating surfaces and modifying the interaction
with the confined fluid. There are several factors that affect the
slip, including wettability, roughness, pressure, temperature or presence
of undissolved gas and nanobubbles\cite{doi:10.1080/00268976.2015.1119899}.
Several studies suggest that surface roughness influences greatly
the boundary conditions of the flow\cite{PhysRevE.74.066311,Zhang2016295,Yen2014,Choi_06,doi:10.1080/00268976.2015.1119899}.
A hierarchical structuration has been reported to reduce notably the
friction\cite{Bixler_12,Zhang_12,PhysRevLett.99.176001}. There is
consensus that hydrophobic coatings reduce friction, due to a weaker
solid-fluid interaction\cite{Liakopoulos2016,:/content/aip/journal/pof2/16/5/10.1063/1.1669400},
but the underlying mechanism remains unclear\cite{PhysRevLett.98.226001}. 

Droplet transport in micro and nanochannels have been thoroughly investigated
in the last decade both theoretically\cite{PhysRevE.74.066311,doi:10.1021/acs.jpcc.5b07951,HE2010126,0953-8984-25-19-195103}
and experimentally\cite{doi:10.1021/la0482406,doi:10.1021/la5004929,doi:10.1021/la504742w,doi:10.1021/la2004744,doi:10.1021/ac3028905}.
Multi-phase flows are important for chemical and biotechnological
applications, because they allow complex phenomena like chemical reactions,
emulsions and interfaces\cite{Shui200735,Rebrov2010} with the potential
of tailoring technological applications. Cao et al\cite{PhysRevE.74.066311}
studied, by Molecular Dynamics (MD) simulations, the flow of a droplet
in coexistence with its gas in a rough nanochannel. They observed
that the slip can be largely influenced by the nano structuration
of the hydrophobic walls. Slug flows are also present in polymer electrolyte
membranes of fuel cells \cite{doi:10.1021/acs.jpcc.5b07951,Zhang01022006,Lu20093445,doi:10.1021/la403057k},
where water has to be eliminated from microchannels at a controlled
rate, to maximize the efficiency of the device. In this context, Fukushima
et al\cite{doi:10.1021/acs.jpcc.5b07951} studied the friction between
a droplet and the confining walls of a microchannel, and its dependence
on the channel width. A review on droplet based microfluidics can
be found in Ref\cite{10.1088/0034-4885/75/1/016601}.

Polymers tethered to a surface through a terminal monomer (polymer
brushes) are good candidates to be used as coating in microchannels,
due to their responsiveness to stimuli (pH, temperature, electric
field, etc) and the ability to modify the rheological conditions\cite{doi:10.1021/ma401537j,Chen201094,doi:10.1021/acs.macromol.7b00450,SMLL:SMLL201402484,POLA:POLA26119,C5SM01962A}.
Polymer brushes present a very wide range of applications, like ph-controlled
nanopores\cite{doi:10.1021/ja8086104}, catalysis in microreactors\cite{doi:10.1021/ja807616z}
and cell adhesion and detachment control\cite{doi:10.1021/la904663m}.
The property of reducing friction of polymer brushes has been reported
theoretically\cite{doi:10.1021/ma8015757,Lee_12,doi:10.1021/ma062875p,C7SM00466D}
and experimentally\cite{doi:10.1021/ma020043v,10.1038/370634a0,doi:10.1021/acs.macromol.5b01267}.
To understand this phenomenon, the boundary conditions of liquids
past fully flexible polymer brushes has been extensively investigated,
and slip lengths of different orders of magnitude have been observed\cite{C5SM02546J,doi:10.1021/acs.macromol.5b02505,Deng_2012,0953-8984-23-18-184105,PhysRevE.80.031608,PhysRevLett.92.018302,0953-8984-20-49-494225,doi:10.1021/ma991796t,doi:10.1021/ma020043v}.
In reference \cite{C5SM02919H}, T. Kreer provides a recent and detailed
review about the lubrication properties of polymer brushes. 

Despite the numerous investigations in the field, little attention
has been given to semiflexible polymer brushes under shear flow. Semiflexible
polymer chains are an adequate coarse-grained model of macromolecules,
whenever the dimensions of the macromolecule do not significantly
exceed its persistence length. Kim et al.\cite{Kim_09} studied via
mean field calculations and computer simulations the response of semiflexible
brushes under shear flow, and found good agreement between both methods
in the high rigidity regime. Deng et al.\cite{Deng_2012} performed
simulations of flow in microchannels coated with semiflexible polymers,
to study glycocalyx fibers for 5 different values of bending rigidity.
Flows are compared keeping constant the channel width, and varying
the stiffness of the fibers. Römer and Fedosov\cite{Roemer_15} extend
the study by Kim et al.\cite{Kim_09} for high grafting densities,
but restricted to stiff polymers. On a recent study, Singh et al.\cite{polym8070254}
studied the trybological behavior of polymer-coated bilayers via MD
simulations, and found that the friction coefficient for semiflexible
polymer brushes was higher than for a fully flexible brush. The comparison
of both types of brushes is made at equal effective width of the channel,
but different grafting densities. In a previous work\cite{C5SM01075F},
we performed a comprehensive study of a nanochannel coated with semiflexible
hydrophobic polymers, filled with a simple liquid. We found that for
low grafting densities $\rho_{g}$, the rigidity of the chains has
a direct influence in the slip length of the flow. To the best of
our knowledge there is no comprehensive study of semiflexible polymer
brushes in shear flow, that analyzes the rheological and chain properties
of the system, covering the whole range of chain stiffness, from fully
flexible polymers to very stiff rod-like polymers. 

\begin{figure}[h]
\begin{centering}
\includegraphics[clip,width=0.95\columnwidth]{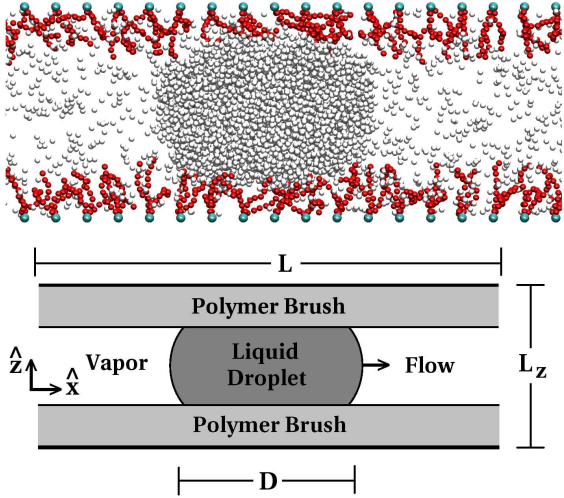}
\par\end{centering}

\protect\caption{\label{fig:Snap}The system studied in this work consists of a slit
like nano-channel, whose walls are separated by a distance $L_{z}$.
Hydrophobic polymers anchored by a terminal bead to the walls coat
the channel, generating a soft substrate. A liquid droplet in coexistence
with its vapor fills the interior of the channel. A constant external
force is applied to every particle to generate net flow of mass. $D$
is the extent of the droplet in the direction of the flow ($\hat{x}$)
and $L$ is the length of the simulation box in the same direction.
Up: snapshot of the simulated system. Down: Schematic representation
of the system.}
\end{figure}

In this work, we perform MD simulations of a liquid-gas two-phase
flow in a slit-like nanochannel, coated by a semiflexible hydrophobic
polymer brushes (see Figure \ref{fig:Snap}). This can be thought
as the flow of a train of droplets through a brush-coated planar nanochannel,
taking into account the periodic boundary conditions. The dependence
of the droplet's velocity is analyzed by changing the distance between
consecutive droplets in the axial direction of the channel, and varying
the rigidity of the polymer chains. We investigate the friction forces
acting on the liquid droplet due to the presence of a gaseous environment
and due to the confining polymer brush, for a wide range of chains'
persistence lengths. We also studied the deformation of the soft substrate
and the conformational changes of the polymers due to the liquid droplet,
and how the brush-liquid interaction varies with bending stiffness.
Lastly, the dynamics of the polymer's free-end is analyzed individually
and collectively. In section \ref{sec:Simulation-Technique} we provide
details of the out-of equilibrium simulation technique, system geometry,
physical conditions and molecular coarse-grained description. In section
\ref{sec:Results} we present the simulation results, focusing on
droplet dynamics and single-chain and brush properties in succesive
subsections. We finish our work with a final discussion and concluding
remarks in section \ref{sec:Concluding-Remarks}.

\section{Simulation Technique\label{sec:Simulation-Technique}}

Coarse-grained molecular dynamics (MD) simulations were performed
at constant temperature, volume and number of particles. All particles
interact with each other via a truncated and shifted Lennard-Jones
potential
\begin{equation}
U(r)=U_{LJ}(r)-U_{LJ}(r_{c}),\,r<r_{c}\,,
\end{equation}
where $r_{c}$ is the cut-off radius, and $r$ is the distance between
particles. Particle pairs whose distance exceeds the interaction cut-off,
do not interact. The parameters of the Lennard-Jonnes potential depend
on the pair of interacting particles 
\begin{equation}
U_{LJ}=4\varepsilon_{\alpha\beta}[(\frac{\sigma}{r})^{12}-(\frac{\sigma}{r})^{6}]\,,
\end{equation}
where the sub-indices $\alpha$ and $\beta$ stand for the different
species. For monomer-monomer and fluid-fluid interactions $\varepsilon$
takes the value of unity, and is the energy unit in the simulations.
For fluid-monomer interactions the energy parameter was set to $\varepsilon_{mf}=1/3$,
which encodes the chemical affinity between the confining polymer
brush and the liquid that flows through the channel. This value of
$\varepsilon_{mf}$ corresponds to a highly hydrophobic substrate,
giving droplet contact angles around $\theta_{c}\sim145\text{º}$.
The cut-off radius for this interactions was set to $r_{c}=2.24\sigma$,
which is twice the distance where the minimum of the potential is
located. This election for the cut-off radius includes the attractive
portion of the potential, allowing for the coexistence of a gaseous
and a liquid phase. For monomer-monomer interactions, the cut-off
radius was set to $r_{c}=1.12\sigma$, to exclude attractive forces
and simulate good solvent conditions. The particle diameter and mass
were set to $\sigma=1$ and $m=1$ for all species, and these parameters
are taken as the distance and mass units respectively. The time unit
can be written in terms of these parameters as $\tau=\sigma\sqrt{m/\varepsilon}$.

The connectivity of the $N=10$ bead polymers chains is provided by
the wide-spread Kremer-Grest model \cite{PhysRevA.33.3628}:
\begin{equation}
U_{FENE}=-\frac{1}{2}kR_{0}^{2}ln\left[1-\left(\frac{r}{R_{0}}\right)^{2}\right]\,,\,r<R_{0}\,,
\end{equation}
where the parameters were set to the usual values $k=30\varepsilon/\sigma^{2}$
and $R_{0}=1.5\sigma$. The finitely extensible nonlinear elastic
(FENE) interaction is applied between consecutive monomers in a chain
to account for the connectivity of each molecule. 

To account for local stiffness of the polymer chains, a harmonic bending
potential was implemented 
\begin{equation}
U_{bend}(\theta)=1/2k_{b}\theta^{2}\,,\label{eq: bending}
\end{equation}
where the angle $\theta$ is defined between two consecutive segments
along the molecule backbone. The orientation of consecutive segments
and the angle $\theta$ are related through the equation
\begin{equation}
cos\theta=\frac{(\boldsymbol{r_{i}}-\boldsymbol{r_{i-1}})\text{·}(\boldsymbol{r_{i+1}}-\boldsymbol{r_{i}})}{|\boldsymbol{r_{i}}-\boldsymbol{r_{i-1}}||\boldsymbol{r_{i+1}}-\boldsymbol{r_{i}}|}\,,
\end{equation}
where $\boldsymbol{r_{i}}$ is the position of particle $i$. 

The parameter $k_{b}$ is the bending constant, which was varied in
this work to modify the rigidity of the polymer brush. To quantify
the rigidity of the chains, we introduce a dimensionless parameter
$l_{p}/l_{c}$. The persistence length $l_{p}$ is the distance over
which segment orientation correlations are lost, and $l_{c}$ is the
contour length of the polymer. When the ratio of this magnitude is
high ($l_{p}/l_{c}\gg1$ ), the polymer behaves like a rigid rod.
On the other hand, a polymer with a low value of this quotient ($l_{p}/l_{c}\ll1$)
behaves like a flexible chain. To estimate the persistence length
of the polymers, we used the exponential decay of the orientational
correlation of the bonds as in Ref\cite{doi:10.1063/1.4971861}: 
\begin{equation}
\langle\cos\theta_{s}\rangle=\exp[-s\text{·}a/l_{p}]\,,
\end{equation}
where $s$ is the distance in monomer units, $a$ is the mean distance
between beads. For consecutive segments ($s=1$), this equation yields
\begin{equation}
l_{p}=-a/\ln\langle\cos\theta_{1}\rangle
\end{equation}

The Velocity Verlet scheme was implemented to integrate the trajectories
of the particles in the simulation, with time step $\delta t=3\text{·}10{}^{-4}$.
Approximately $6\text{·}10^{7}$ steps were performed for each simulation,
including $\sim10^{6}$ relaxation steps. To maintain a constant value
of temperature during the simulations, the dissipative particle dynamics
(DPD) algorithm was applied\cite{Hoogerbrugge_92,Espanol_95,Soddemann_03}.
The equations of motion are modified as follows:
\begin{equation}
\boldsymbol{F}=\boldsymbol{F_{C}}+\boldsymbol{F_{R}}+\boldsymbol{F_{D}}\,,
\end{equation}
where $\boldsymbol{F_{C}}$ is the sum of the conservative forces,
$\boldsymbol{F_{R}}$ is a random force and $\boldsymbol{F_{D}}$
is the dissipative force. All forces are applied in pairs, abiding
Newton's third Law, therefore conserving momentum locally. For each
interacting pair, a random force is calculated as follows:
\begin{equation}
\boldsymbol{F_{R}}=\zeta\omega_{R}(r)\eta\hat{\boldsymbol{r}}\,,
\end{equation}
where $r$ is the distance between particles, $\hat{r}$ is the unity
vector joining a given pair of particles, $\zeta$ is a parameter
that modifies the strength of the force, $\eta$ is a random number
generated for each occurrence, and the weight $\omega_{R}$ is a function
of the distance between particles. The dissipative Force is:

\begin{equation}
\boldsymbol{F_{D}}=-\gamma\omega_{D}(r)(\hat{\boldsymbol{r}}\text{·}\boldsymbol{v})\hat{\boldsymbol{r}}\,,
\end{equation}
where $\gamma$ is the friction constant, $\boldsymbol{v}$ is the
difference of velocities of the particles, and $\omega_{D}$ is also
a weight function. The parameters and functions were chosen to fulfill
the fluctuation-dissipation theorem, which happens for the relations:
$\zeta^{2}=2k_{B}T\gamma$ and $\omega_{R}^{2}(r)=\omega_{D}(r)$.
The usual choice for these weight functions was made\cite{Pastorino_07},
and the friction constant was set to $\gamma=0.5.$ All simulations
were performed at a fixed temperature value $T=0.8\varepsilon/k_{B}$.
At this temperature, the Lennard-Jones fluid separates into two phases:
a liquid phase of density $\rho_{l}=0.69\sigma^{-3}$ , and a vapor
phase of density $\rho_{v}=0.03\sigma^{-3}$. 

Periodic boundary conditions were applied in the directions parallel
to the walls ($\hat{x}$ and $\hat{y}$), while in the remaining direction
($\hat{z}$) a purely repulsive and smooth 9-3 potential $U_{wall}$
was implemented to prevent the particles from escaping the simulation
box and to provide a planar channel geometry for the system.
\begin{equation}
U_{wall}(z)=A_{w}\left[\left(\frac{\sigma}{z}\right)^{9}+\left(\frac{\sigma}{z}\right)^{3}\right]\,,
\end{equation}
with $A_{w}=3.2$. The simulations box dimensions are $L_{x}\times L_{y}$$=13.42\sigma\times L_{z}$,
where the length of the simulation box in the flow direction was varied
between $35\sigma<L_{x}<900\sigma$, and wall to wall distances were
adjusted in the range $30\sigma<L_{z}<40\sigma$. 

The polymer brush is composed of linear polymer chains of $N=10$
beads each, with a terminal monomer fixed to a wall. The grafting
density (number of chains per unit area) was set to $\rho_{g}=0.05\sigma^{-2}$
for all simulations, and the grafted sites were arranged in an ordered
square lattice of parameter $a_{l}=1/\text{\ensuremath{\sqrt{\rho_{g}}}}=4.47\sigma$.
A uniform distribution of grafting sites was chosen to avoid large
inhomogeneities in the polymer density profile that can appear in
small system with low $\rho_{g}$ when the chain heads are randomly
scattered. We think that this way of distributing the chains is more
representative than the random like, because the latter can produce
singular physical phenomena that depend on the particular arrangement
of the polymers, specially for the relatively small samples used in
the simulations. 

To favor a perpendicular orientation of the polymers with respect
to the channel walls, a virtual bead is added to each polymer chain
in the bending force calculation. This virtual bead is placed below
the grafted monomer, and when the bending force is calculated, the
bond between this bead and the grafted end-bead is used to calculate
a bending force on the second bead of the chain. This force tends
to align the first mobile monomers of the chain in the direction perpendicular
to the wall, and to induce a stretching of the polymers towards the
center of the channel.

An external force is applied to all particles in the system during
the simulations to create a flow, and take the system out of equilibrium.
This body force is constant and is applied in a direction parallel
to the walls of the channel ($\hat{x}$). The value $f_{ext}=0.002\varepsilon/\sigma$
was chosen to obtain particle velocities in the range $0.1\sigma/\tau\,<v<0.5\sigma/\tau$
which are high enough to extract significant data in a reasonable
computation time, and is low enough to be in a near-equilibrium regime.

\section{{\large{}Results\label{sec:Results}}}

\subsection{Liquid - Polymer Brush Interface}

In this section we analyze the interaction between semiflexible polymers
grafted to the interior walls of the nanochannel and a liquid drop
flowing through it. The polymer-liquid interaction potential parameters
were chosen such that the droplet is in super-hydrophobic regime (
Casie-Baxter state). We focus our attention on the influence of the
polymer rigidity on the droplet flow, which is studied by varying
the bending constant $k_{b}$ (see Eq. \ref{eq: bending}) in the
range $0\leq k_{b}\leq160\varepsilon$ or, in terms of persistence
length over contour length, in the range $0.1\leq l_{p}/l_{c}\leq20$.
The length of the channel in the $\hat{x}$ direction was set to $L=107\sigma$
for this set of simulations, while the extension of the droplet in
the same direction is approximately $D=30\sigma$. The droplet trajectory
is tracked, and the mean velocity is calculated performing a linear
fit. Details of the non-trivial tracking algorithm for the center
of mass of the droplet are presented in the Supporting Information.
It is not straight forward to compare systems with different degrees
of local stiffness, because the polymer extension depends strongly
on $k_{b}$. A hard bending potential will give rise to a polymer
brush of larger height, consequently decreasing the effective width
of the channel and compressing the droplet. We kept constant the effective
channel width in order to isolate the influence of bending rigidity
on the flow, by comparing the shape of the droplets inside the channels.
To maintain the shape of the droplet similar for all the studied bending
rigidities, it is necessary to adjust the distance between walls ($L_{z}$).
The case of totally flexible chains ($k_{b}=0)$ was taken as reference
to compare the two-dimensional density profiles of the droplets. For
each system (i.e. for each value of bending constant $k_{b}$), various
simulations were performed for different channel widths ($L_{z}$),
to find the droplet shape that most resembles the reference case.
Details of this procedure are provided in the Supporting Information.

\begin{figure}[h]
\centering{}\includegraphics[clip,width=0.95\columnwidth]{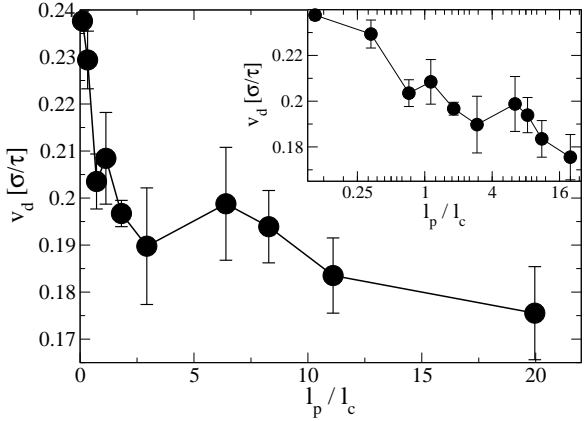}\protect\caption{\label{fig:veldroplet_vs_lpersistence}Droplet velocity versus persistence
length of the polymers. Two different regimes can be observed: for
persistence length smaller than the contour length, the velocity diminishes
rapidly. For persistence lengths greater than the contour length,
the droplet's velocity decreases with persistence length, but at a
much more slower rate. The inset shows the droplet velocity plotted
against the persistence length in a semi-log graph to highlight the
behavior at low persistence lengths. }
\end{figure}

In Figure \ref{fig:veldroplet_vs_lpersistence} the droplet velocity
is plotted against bending rigidity. The large error bars are due
to the uncertainty to determine the channel width ($L_{z}$), to obtain
similar droplets for all bending rigidities. To quantify the rigidity
of the polymers, we utilized the dimensionless parameter $l_{p}/l_{c}$,
where $l_{p}$ is the persistence length and $l_{c}$ is the contour
length, as explained in section \ref{sec:Simulation-Technique}. For
persistence lengths ($l_{p}$) smaller than the contour length ($l_{c}$),
the velocity of the droplet decays rapidly with increasing stiffness,
while for $l_{p}/l_{c}>1$ the velocity of the droplet decreases slowlyer
with rigidity. The general tendency is that increasing the rigidity
of the polymer, reduces the droplet velocity. 

\begin{figure}
\centering{}\includegraphics[width=0.95\columnwidth]{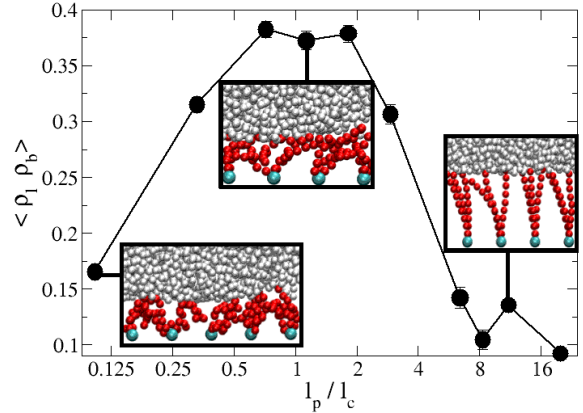}\protect\caption{\label{fig: Brush_Liq_Interactions}Average of the product of liquid
and brush densities. It is proportional to the number of interactions
between brush and droplet. Insets: morphology of the brush polymer
layer for fully flexible polymers, $l_{p}/l_{c}=0.1$ (left) medium
stiffness, $l_{p}/l_{c}=1.1$ (center) and very high stiffness, $l_{p}/l_{c}=11$
(right). The vapor particles are not shown in order to have a clear
visualization of the polymer chains. }
\end{figure}

To gain insight into the friction dependence on chain stiffness, the
average of the product of the density profiles between liquid and
brush is plotted against rigidity in Figure \ref{fig: Brush_Liq_Interactions}.
This quantity is proportional to the number of brush-liquid interactions\cite{pastorino:064902}.
It is interesting to observe the non-monotonic dependence of the number
of monomer-droplet particle interactions with the polymer rigidity
(see Figure \ref{fig: Brush_Liq_Interactions}). As shown in the left
inset of Figure \ref{fig: Brush_Liq_Interactions}, the fully flexible
chains ( $l_{p}/l_{c}\ll1$) are in a mushroom regime. The polymers
wrap themselves to maximize conformational entropy and do not penetrate
the liquid phase so frequently. More rigid polymers $(l_{p}/l_{c}\simeq1)$
adopt a banana-like shape, due to the drag force exerted by the liquid
droplet. This exposes more beads of each chain to the droplet, thus
increasing the number of brush-liquid interactions. On the other hand,
highly rigid polymers $(l_{p}/l_{c}\gg1)$ tend to extend themselves
in the direction perpendicular to the substrate. The low inclination
angle of these stiff polymers hinders the non-terminal beads from
reaching the liquid phase. Only the free end-bead monomers of each
chain are in contact with the droplet, thus reducing the number of
collisions. It is important to note that the hydrophobicity between
liquid and polymers warrants a Cassie-Baxter regime in all cases,
but with a significantly different structure of the polymer brush.
The number of monomer-liquid interactions alone cannot account for
the dependence of the droplet velocity on bending rigidity, observed
in Fig. \ref{fig:veldroplet_vs_lpersistence}. This implies that there
is another property of the system that varies with $l_{p}$ and explains
the decrease in the flow.

\begin{figure}
\centering{}\includegraphics[clip,width=0.95\columnwidth]{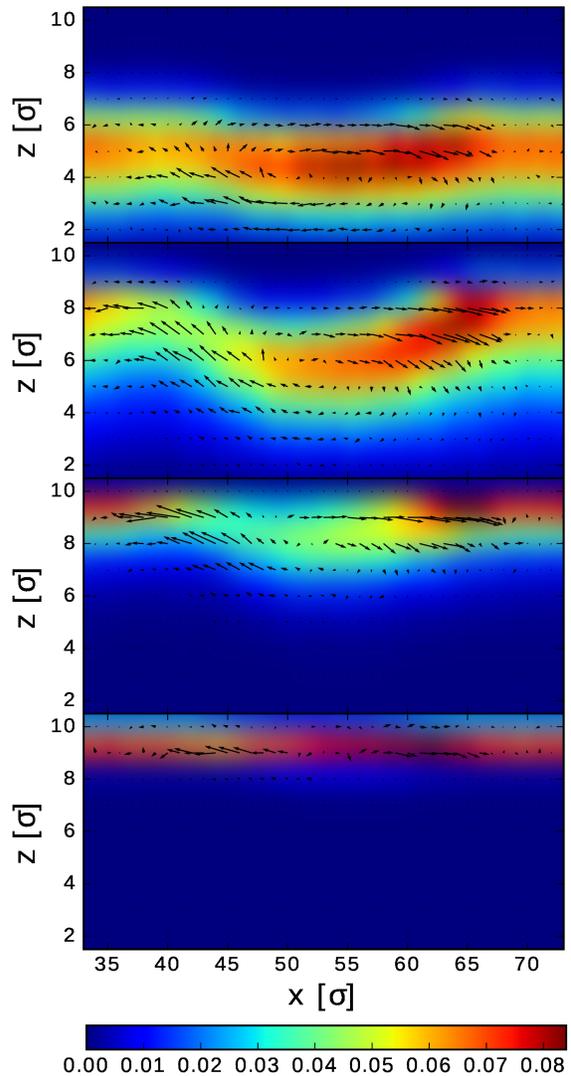}\protect\caption{\label{fig:Rend_2D_histo}The polymer's end-bead distribution is shown
as a color map. Green and yellow zones represent high presence of
terminal monomers, while blue zones represent low density of terminal
beads. The red arrows, composing a vector field, represent the mean
momentum of the end-beads of the polymers. These graphs are presented
for various rigidities, from top to bottom: $l_{p}/l_{c}=0.1;\,0.72;\,2.9;\,8.3$.}
\end{figure}

The 2-dimensional terminal-bead distribution density (see Figure \ref{fig:Rend_2D_histo})
shows how the brush is deformed by the liquid droplet, for different
values of bending rigidity. In every case the brush near the advancing
end of the droplet (right) is denser than near the back (left). This
is due to the liquid particles colliding with the polymers, pushing
them forward, squeezing the monomers near the advancing end of the
droplet. It is also observed that the brush composed of slightly rigid
chains ($l_{p}/l_{c}=0.33$) suffers a stronger deformation by the
droplet, than the brush composed of fully flexible polymers ($l_{p}/l_{c}=0.1$).
The latter is denser, hence there are more excluded volume interactions
impeding the compression of the brush. For very rigid chains ($l_{p}/l_{c}\geq6.4$),
the end-bead distribution narrows, and its mean value gets closer
to the center of the channel, due to the polymer elongation in $\hat{z}$.

To analyze the collective motion of the brush, the terminal-beads
momentum vector field is plotted over the density profile (black arrows
in Figure \ref{fig:Rend_2D_histo}). From this graph it is possible
to extract qualitative information about the dynamics of the polymer
brush. First, it can be noted that for all degrees of stiffness, the
polymers that are in contact with the advancing end of the droplet
(right side in Fig. \ref{fig:Rend_2D_histo}), tend to have a velocity
in the direction of the flow and towards the wall. The droplet collides
with the polymers in the flow direction, and compresses the brush
towards the channel walls. Second, near the receding contact angle
of the droplet (left region of the graphs in Fig. \ref{fig:Rend_2D_histo}),
the brush has a mean velocity towards the center of the channel, in
the $\hat{z}$ direction. The liquid applies a force on the brush
perpendicularly to the grafting plane $(\hat{z})$, pressing the end
bead of the chains near the wall. When the droplet passes, the force
on the chains is no longer applied, and the free ends rise towards
the center of the channel.

In third place, for flexible polymers, the layer in contact with the
liquid droplet has a positive mean velocity, while the terminal monomers
near the wall have a mean velocity in the direction opposite to the
flow. This collective dynamics resembles the motion of a treadmill
belt. A similar behavior has been reported for flexible polymer brushes
under shear: a velocity profile which is negative in the interior
of the brush layer , and positive near the brush-liquid interface
\cite{Mueller_08b,:/content/aip/journal/jcp/140/1/10.1063/1.4851195}.
This velocity profile is the result of individual polymers performing
a cyclic motion, which consist of polymers stretching in the direction
of the flow after spontaneous excursions inside the liquid driven
by thermal fluctuations. Afterwards the chains retracts to the grafted
site closer to the wall, to maximize their configurational entropy.
The bending potential hinders the cyclic motion of the polymer chains,
because it competes with the configurational entropy, stretching the
chains and inhibiting the retraction movement\cite{C5SM01075F}. Increasing
the persistence length of the polymers, flattens the velocity profile
of the polymer brush.

To exhibit the influence of the brush dynamics on the flow observed
in Figure \ref{fig:Rend_2D_histo}, we examined the velocity of the
brush in contact with the liquid. As mentioned before, the region
of space where brush and droplet interact will be given by the overlap
of the liquid density $\rho_{l}(\boldsymbol{r})$ and brush density
$\rho_{b}(\boldsymbol{r})$. Zones of high brush-droplet interaction
are portrayed by a high value of the product of the liquid and brush
densities $\rho_{l}(\boldsymbol{r})\text{·}\rho_{b}(\boldsymbol{r})$,
while a null value of $\rho_{l}(\boldsymbol{r})\text{·}\rho_{b}(\boldsymbol{r})$
indicates vanishing brush-liquid interaction. We define the brush
boundary layer velocity as:
\begin{equation}
v_{b}^{(BL)}\equiv\frac{\int d\boldsymbol{r}\rho_{l}(\boldsymbol{r})\rho_{b}(\boldsymbol{r})v_{b}(\boldsymbol{r})}{\int d\boldsymbol{r}\rho_{l}(\boldsymbol{r})\rho_{b}(\boldsymbol{r})}\,,
\end{equation}
where $v_{b}(\boldsymbol{r})$ is the velocity field of the monomers
composing the polymer brush, $\rho_{b}(\boldsymbol{r})$ and $\rho_{l}(\boldsymbol{r})$
are the brush and liquid number densities, respectively. This magnitude
($v_{b}^{(BL)}$) is the weighted average brush velocity, where the
weight is given by the product of the brush and liquid densities.
To take into account only the liquid phase, and leave the vapor contribution
out, we set the liquid density to $\rho_{l}(\boldsymbol{r})=0$, if
the fluid density is below the average of the equilibrium liquid and
gas densities. Analogously we define the liquid boundary layer velocity
as:

\begin{equation}
v_{l}^{(BL)}\equiv\frac{\int d\boldsymbol{r}\rho_{l}(\boldsymbol{r})\rho_{b}(\boldsymbol{r})v_{l}(\boldsymbol{r})}{\int d\boldsymbol{r}\rho_{l}(\boldsymbol{r})\rho_{b}(\boldsymbol{r})}\,,
\end{equation}
which can be interpreted as the mean velocity of the liquid in contact
with the polymer brush. $v_{l}(\boldsymbol{r})$ is the velocity field
of the liquid phase.

\begin{figure}
\centering{}\includegraphics[clip,width=0.95\columnwidth]{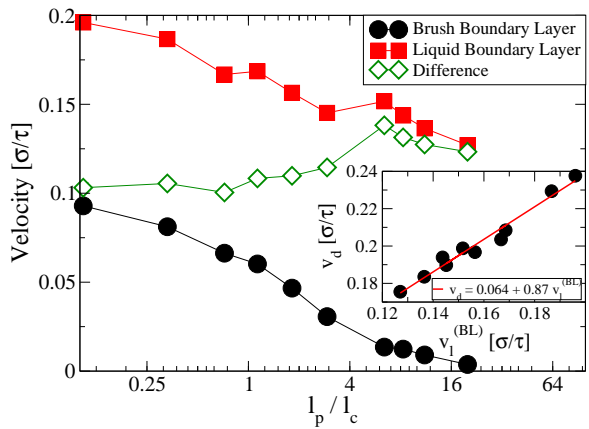}\protect\caption{\label{fig:SlipVel_vs_rigidity}Velocity of the brush (black circles)
and liquid (red squares) in the brush-liquid interface are plotted
as a function of persistence length. The difference between these
velocities, shows how the liquid boundary layer passes over the brush
boundary layer (open green diamonds). Inset: droplet velocity versus
the liquid boundary layer velocity. The liquid boundary layer velocity
is closely related to the effective slip boundary condition for the
flow.}
\end{figure}

The brush boundary layer velocity ($v_{b}^{(BL)}$) and the liquid
boundary layer ($v_{l}^{(BL)}$) velocity are plotted against persistence
length in Figure \ref{fig:SlipVel_vs_rigidity}. $v_{l}^{(BL)}$ is
closely related to the effective slip boundary condition of the flow
of the simple liquid inside the channel, used in hydrodynamic calculations.
Assuming a poiseuille flow near the center of the liquid phase, the
droplet velocity should depend linearly on the slip velocity $v_{s}$,
with a slope equal to one. In the inset of Figure \ref{fig:SlipVel_vs_rigidity},
the droplet velocity ($v_{d}$) is plotted against $v_{l}^{(BL)}$.
It can be observed that $v_{d}$ varies linearly with the liquid boundary
layer velocity $v_{l}^{(BL)}$, with a slope of $0.87$. Therefore,
we can roughly identify $v_{l}^{(BL)}$ as the effective slip boundary
condition imposed by the polymer brush on the droplet flow.

It is interesting to analyze the brush boundary layer velocity $v_{b}^{(BL)}$
as rigidity increases (see black circles in Figure \ref{fig:SlipVel_vs_rigidity}).
For flexible polymers ($l_{p}/l_{c}\ll1$), $v_{b}^{(BL)}$ has the
same order of magnitude as the droplet velocity, while for stiff chains
($l_{p}/l_{c}\gg1$), $v_{b}^{(BL)}$ drops to zero. The overall average
brush velocity is null, because the chains are grafted to the static
walls. The dynamics of the flexible polymers in the brush reduces
the friction between liquid and substrate, by adopting a positive
velocity near the liquid and a negative velocity near the wall (see
Figure \ref{fig:Rend_2D_histo}). As chain stiffness increases, the
cyclic motion of the polymers is hindered, because the bending rigidity
impedes the wrap and recoil movement\cite{C5SM01075F}. This ``treadmill
belt'' like dynamical behavior of the chains affects heavily the
final velocity of the droplet. The correlation coefficient between
the droplet velocity and the brush boundary layer velocity is $corr(v_{d},v_{b}^{(BL)})=0.91$.
To the best of our knowledge, there is no previous quantitative work
that shows that the internal dynamics of a polymer brush can affect
the droplet flow.

We also define the $\delta v$ as the difference between the liquid
boundary layer velocity $v_{l}^{(BL)}$ and the brush boundary layer
velocity $v_{b}^{(BL)}$. This magnitude measures how the liquid boundary
layer passes the brush boundary layer. $\delta v$ maintains a fairly
constant value in the range $l_{p}/l_{c}<4$, and for $l_{p}/l_{c}>4$
there is a sudden increase. This change in the behavior of $\delta v$
coincides with the decrease of the number of polymer-liquid interactions
(see Figure \ref{fig: Brush_Liq_Interactions}). We think that the
liquid boundary layer velocity relative to the brush boundary layer
increases with rigidity, due to the decrease of polymer-liquid interactions. 

To summarize the effect of polymer's bending rigidity on flow properties,
we can distinguish two factors. First, the internal dynamics of the
polymers facilitates the liquid flow for flexible polymers. This effect
is hindered as the chain's bending rigidity increases. Secondly, the
way in which the closest layers of liquid and brush slide past each
other. The sliding of liquid relative to the brush is enhanced for
large persistence lengths $(l_{p}/l_{c}>1)$. These two effects compete
with each other, and the overall brush motion seems to have a more
determinant role on the droplet's velocity.

\subsection{Brush and chain properties \label{sub:Brush-and-chain}}

\begin{figure}
\begin{centering}
\includegraphics[clip,width=0.95\columnwidth]{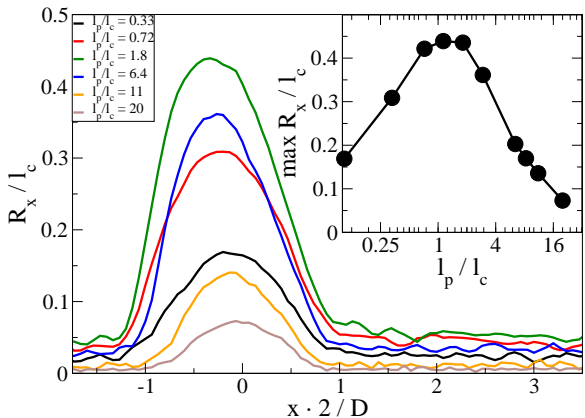}
\par\end{centering}

\centering{}\protect\caption{\label{fig:Rx_vs_rigidity}End-to-end brush vector component in the
flow direction ($\hat{x}$) versus position ($x$), normalized by
the droplet size ($D$). Data is analyzed in the droplet frame of
reference, where the center of mass of the droplet is always on $x=0.$
Inset: Maximum inclination of the end-to-end brush vector component
in the direction of the flow ($\hat{x}$) versus rigidity ($l_{p}/l_{c}$).
A non-monotonic behavior can be observed as the rigidity of the polymers
increases (see text). }
\end{figure}

In the previous section we have shown that the dynamics of the polymer
chains affects the rheological properties inside the nano-channel.
The influence of brush deformations on friction forces has also been
reported in a recent study\cite{doi:10.1021/acs.langmuir.7b00217}.
Therefore, it is worth studying the conformation and dynamical behavior
of the polymer brush.

To quantify the deformation of the brush in the flow direction ($\hat{x}$),
the $\hat{x}$ component of the end-to-end vector ($\boldsymbol{R}$)
is plotted against the $x$ coordinate for various degrees of rigidity
in Figure \ref{fig:Rx_vs_rigidity}. It can be observed that near
the center of mass of the droplet ($x=0$), the inclination of the
chains reaches a maximum in all cases. This is due to droplet passing
over the polymer brush, elongating the chains in the direction of
the flow. The graph shows that the chain inclination profiles depend
on the rigidity in a non-monotonic way. To expose this dependence,
the maximum inclination is plotted versus the persistence length of
the chain (see Inset of Figure \ref{fig:Rx_vs_rigidity}). Flexible
chains ($l_{p}/l_{c}<1$) tend to wrap themselves, to maximize their
configurational entropy. Increasing the bending rigidity induces a
coherent stretching along the chain, increasing the distance between
the terminal monomers of the chains. When the persistence length ($l_{p}$)
is similar to the contour length ($l_{c}$) the displacement of the
chains reaches a maximum ($l_{p}/l_{c}\simeq1)$, and for $l_{p}/l_{c}>1$
it decreases. The first bond of the polymer chains is oriented perpendicular
to the wall (see section \ref{sec:Simulation-Technique}), which favors
a vertical direction of elongation. For $l_{p}/l_{c}\gg1$ the angles
between consecutive bonds are strongly correlated through the whole
chain, thus the orientation of the first bond (in $\hat{z}$) endures
until the last monomer. In this regime, bending forces hinder the
bonds to reach large angles between consecutive monomers, therefore
the elongation of the chains takes place mainly in the direction perpendicular
to the wall and not in the parallel direction.

\begin{figure}
\centering{}\includegraphics[width=0.95\columnwidth]{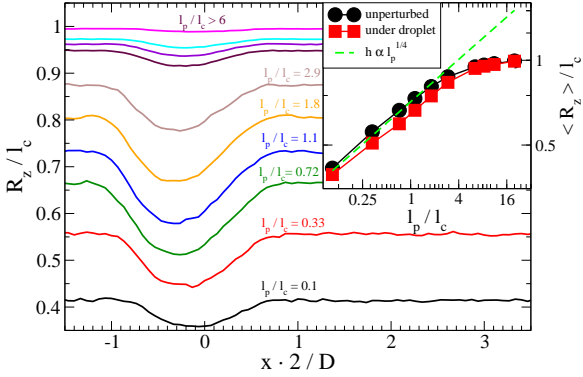}\protect\caption{\label{fig:Rz_vs_x}Profile of the end-to-end brush vector component
in the direction perpendicular to the walls ($\hat{z}$) versus $x$
coordinate, in the droplet frame of reference . The horizontal axis
is normalized, dividing by the length of the droplet in the direction
of the droplet velocity. Inset: Mean value of $R_{z}$ unperturbed
(black circles) and under the droplet (red squares) in a log-log plot.
The dashed line is the scaling law predicted by Mean Field theory
\cite{BIRSHTEIN19832165}.}
\end{figure}

In Figure \ref{fig:Rz_vs_x}, the profile of the vertical component
of the end-to-end vector ($R_{z}$) is presented for every studied
bending rigidity. The height of the brush increases monotonically
with the bending rigidity, and in all cases the profile of $R_{z}$
presents a minimum near the droplet's center of mass. The droplet
applies pressure on the polymer brush, decreasing its height locally.
To have a better visualization of the the behavior of the polymers,
the mean height of the polymers in the presence of the droplet $|x|<D_{{\rm droplet}}$
and far away from it $|x|>D_{{\rm droplet}}$ are plotted against
rigidity, in the inset of Figure \ref{fig:Rz_vs_x}. The height of
the brush, unaffected by the droplet, scales as $h\,\propto\,l_{p}^{1/4}$
with persistence length (dashed line), and for $l_{p}/l_{c}\gg1$
the increase rate diminishes due to finite extension effects. The
scaling law of brush height with persistence length was first proposed
by Birshtein and Zhulina\cite{BIRSHTEIN19832165} for semiflexible
polymer brushes in mushroom regime (see Table I). A similar behavior
for the unperturbed brush height was also observed by Kim et al.\cite{Kim_09},
who performed Lattice-Boltzmann and Brownian-Dynamics simulations
(see their Figure 3b), and by Milchev and Binder\cite{Milchev_14b}
by MD simulations (see their Figure 2b). The vertical extension of
the polymers under the droplet (red squares) follows basically the
same behavior, but with small deviations. These are due to the different
compressibilities of the brush, which vary with the bending rigidity.

\begin{figure}
\centering{}\includegraphics[clip,width=0.95\columnwidth]{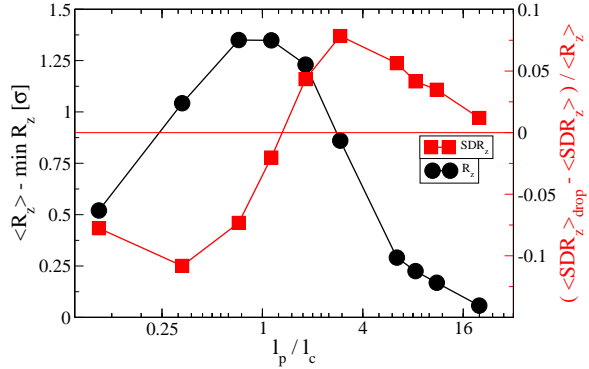}\protect\caption{\label{fig:SDRz_vs_rigidity}Difference between vertical stretching
of the polymers under the droplet and far away from it (black circles).
This magnitude quantifies the deformations of the brush in the directions
perpendicular to the wall, due to the presence of the droplet. In
red squares, it is shown the mean displacement of the end-to-end vector
under the droplet with respect to the unperturbed displacement in
the direction perpendicular to the wall. This magnitude gives an idea
of the typical movement of the polymer's end-bead, and how it changes
in the presence of the droplet. For low rigidities the displacement
is hindered, while for high rigidities the end-bead movement is enhanced. }
\end{figure}

To quantify the perturbation of the brush due to the presence of the
liquid, in Figure \ref{fig:Rz_vs_x}, we calculated the mean value
of the unperturbed brush height minus the minimum of the $R_{z}(x)$
profile. This is plotted as a function of bending rigidity in Fig.
\ref{fig:SDRz_vs_rigidity} (black circles). The deformation of the
brush, caused by the presence of the droplet, is maximum for persistence
lengths similar to the contour lengths ($l_{p}/l_{c}\simeq1$). Fully
flexible polymer-brushes ($l_{p}/l_{c}\ll1$) are more dense than
brushes composed of semi-flexible chains, and the excluded volume
interactions between monomers hinders the compression of the brush.
In the other extreme, very rigid polymers ($l_{p}/l_{c}\gg1$) are
also difficult to compress vertically, because chains are strongly
stretched in the direction perpendicular to the wall, to reduce the
elastic energy. A compression towards the wall requires a high energy
to compensate the increase in elastic energy. In between these regimes
($l_{p}/l_{c}\simeq1$), the brush deformation reaches a maximum.
In this case the segments in each chain elongate coherently in a defined
direction, and the polymers adopt a banana-shape configuration, when
compressed by the droplet. This particular conformation allows for
an easier compression than the fully flexible chains, because in this
range ($l_{p}/l_{c}\simeq1$), the bending potential is softer than
the excluded volume potential. 

The dynamics of the polymer chains change qualitatively with the bending
constant $k_{b}$. The standard deviation of the end-to-end vector
distribution quantifies the movement of the free end of each chain
with respect to its mean position. By calculating the standard deviation
of $R_{z}$ for polymers under the droplet and far away from it, it
is possible to shed light on the influence of the liquid in the dynamics
of the chains. In Figure \ref{fig:SDRz_vs_rigidity} the difference
between the mobility of the terminal bead in the presence of the liquid
and far away from the droplet is plotted against stiffness (red squares).
A negative value means that the presence of the droplet hinders the
motion of the chains, while a positive value means that the interaction
droplet-brush enhances chain mobility. For flexible polymers, the
space that the terminal bead can explore is limited by the contour
length of the polymer, due to the connectivity of the molecule. In
the presence of a droplet, the available space to explore is reduced,
as the chain does not penetrate significantly in the liquid phase.
This leads to a reduction in the mobility of the polymers. Stiffer
polymers extend in the direction perpendicular to the wall ($\hat{z}$),
and in the presence of a liquid droplet they bend, forming an arc.
For $l_{p}/l_{c}>1.4$, the bending potential is hard enough such
that the curved chains elongate frequently in $\hat{z}$, and penetrate
in the droplet. This leads to an increase in the chain mobility in
the direction perpendicular to the wall.

\begin{figure}
\centering{}\includegraphics[clip,width=0.95\columnwidth]{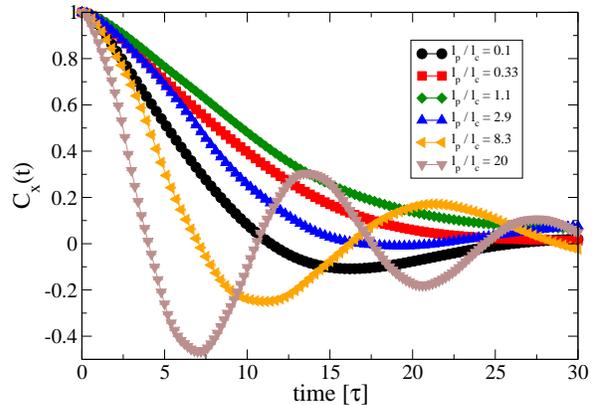}\protect\caption{\label{fig:Time-autocorrelation-function}Time autocorrelation function
of the $\hat{x}$ component of the end to end vector for different
values of rigidity. The values of relative stiffness $l_{p}/l_{c}$
are shown in the legend. }
\end{figure}

To gain further insight in the mean individual polymer dynamics, the
time autocorrelation function of the end-to-end vector in the direction
of the flow was calculated for several values of stiffness (see Figure
\ref{fig:Time-autocorrelation-function}). This quantity is defined
as 
\begin{equation}
C_{x}(t)=\frac{\langle R_{x}(0)\text{·}R_{x}(t)\rangle-\langle R_{x}^{2}\rangle}{\langle R_{x}\rangle^{2}-\langle R_{x}^{2}\rangle}\,,
\end{equation}
where $\langle f\rangle$ denotes the time average of $f$. The effect
of the bending forces is evidenced in the oscillatory motion of the
stiffest chains $l_{p}/l_{c}\gg1.$ As expected, the more rigid the
bending potential, the higher the oscillation frequency. For $l_{p}/l_{c}<3$
no oscillatory motion can be observed. A decay of $C_{x}(t)\rightarrow0$
can be observed in all cases, which corresponds to the decorrelation
of the trajectory of the free end of the polymers.

\subsection{Fluid Dynamics\label{sub:Droplet-Dynamics}}

\begin{figure}[h]
\centering{}\includegraphics[width=0.95\columnwidth]{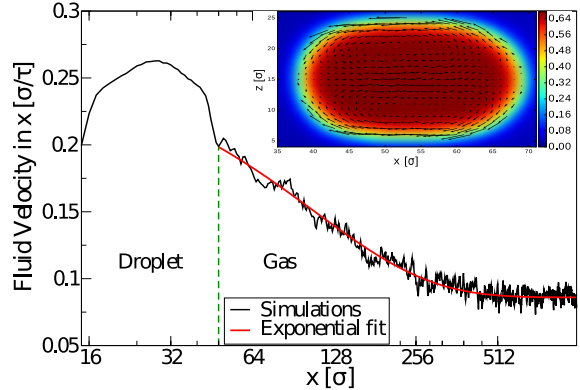}\protect\caption{\label{fig:Fluid_vel_vs_x}Fluid velocity in the direction of the
flow ($\hat{x}$) as a function of the x coordinate (in log scale).
The droplet ($x<48\sigma$) travels at a higher velocity than the
vapor phase ($x>48\sigma$). It is also possible to observe that gas
velocity decays exponentially with x. Fitting the fluid velocity by
an exponential function, a relaxation length ($l_{rel}$) was obtained
for the gas $l_{rel}=107\sigma$. The terminal velocity of the gas
$v_{\infty}$(far away from the droplet) coincides with the simulations
done in the absence of a liquid phase. Inset: 2-dimensional density
profile of the droplet, with the liquid velocity field overlayed.
The velocities are calculated in the frame of reference of the center
of mass of the droplet.}
\end{figure}

In this section we study the properties of the two-phase liquid and
gas flows in the nanochannel. The fluid velocity at $z=L_{z}/2$ is
plotted as a function of the $x$ coordinate in Figure \ref{fig:Fluid_vel_vs_x}
for a long channel ($L=900\sigma$). The droplet, located in $x<48\sigma$,
has a higher velocity than the gas ($x>48\sigma$). The velocity of
the gas near the droplet matches the velocity of the liquid, and then
decays exponentially to the terminal velocity $v_{\infty}=0.85\sigma/\tau$.
By fitting the velocity profile of the gas with an exponential function,
the characteristic relaxation length of the gas was found to be $l_{rel}=107\sigma$. 

The internal velocity field of the droplet is shown in the inset of
Figure \ref{fig:Fluid_vel_vs_x}. A recirculation flow is observed
with two stagnation points near the brush-liquid interface. Similar
vector fields were reported by Günther et al.\cite{doi:10.1021/la0482406}
for experiments on liquid segments in microchannels, in coexistence
with gas (see their Figure 5b). This internal flow affects the droplet
velocity, due to the viscous dissipation. 

\begin{figure}[h]
\centering{}\includegraphics[clip,width=0.95\columnwidth]{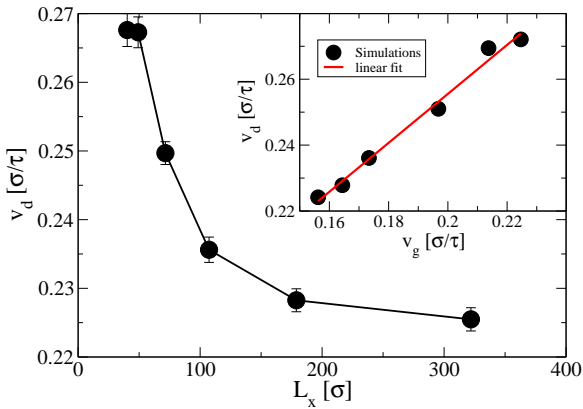}\protect\caption{\label{dropVel_vs_L_kb0}Droplet velocity ($v_{d}$) as a function
of the channel length ($L$). Extending the channel length $L$ is
equivalent to increasing the distance between consecutive droplets,
due to the periodic boundary conditions applied in the simulations.
This graphic shows that the closer the droplets are to each other,
the faster they move through the channel. Inset: Droplet velocity
(black circles) plotted against surrounding gas velocity. Each point
corresponds to a different channel length. The red continuous line
is the linear regression of the data. }
\end{figure}

The difference between the gas and liquid velocities implies that
the droplet moves relative to the surrounding gas. Therefore, the
liquid suffers a drag force applied by the coexisting vapor. If the
channel is not long enough to allow for the gas phase to relax ($L\leq2l_{rel}$)
and adopt its terminal velocity, then the mean gas velocity will be
closer to the velocity of the droplet. Because the gas friction force
is proportional to the velocity difference between vapor and droplet,
in a short channel the friction will be lower, thus producing a faster
moving droplet, as seen in Figure \ref{dropVel_vs_L_kb0}. This can
be also observed by noting that due to the periodic boundary conditions
in $\hat{x}$ , the channel length $L$ is equal to the distance between
consecutive droplets in an infinitely long channel. The system represents
a train of droplets, each at a fixed distance, given by $L$. If the
distance between adjacent droplets is large, then the gas between
them relaxes and adopts the terminal gas velocity $v_{\infty}$. This
will enhance the velocity difference between gas and droplet, which
is proportional to the friction force, thus slowing the droplets down. 

To portray the gas-liquid interaction, a set of simulations were performed
only varying the length of the simulation box in the $x$ direction,
$L$. The bending constant was fixed to $k_{b}=0$, and the number
of liquid particles in the droplet was set to ($n_{d}\simeq4200$).
In Figure \ref{dropVel_vs_L_kb0}, the velocity of the droplet is
plotted against channel length $L$. Despite maintaining the number
of particles in the droplet constant ($n_{d}\simeq4200$), and applying
the same total external body force on them, the droplet velocity varies
significantly ($\sim16\%$ ) with $L$.

To verify the model of drag friction between gas and liquid ($f_{g}=-\gamma_{{\rm gas}}(v_{d}-v_{{\rm gas}})$),
the equation of motion of the droplet can be examined. The dissipative
forces are proportional to the droplet velocity\cite{0953-8984-25-19-195103,0953-8984-23-18-184105,doi:10.1021/acs.jpcc.5b07951}
$f_{{\rm dis}}=-\alpha\eta v_{d}$, where $\eta$ is the liquid viscosity,
and $\alpha$ is a constant of the order of unity. The total driving
force is the sum of the body forces applied to each particle in the
droplet $F_{{\rm ext}}=n_{{\rm drop}}f_{{\rm ext}}$. The sum of all
forces acting on the droplet must be zero on average, because the
trajectory of the droplet describes a uniform linear motion:

\begin{equation}
0=n_{{\rm drop}}f_{{\rm ext}}-\gamma_{{\rm gas}}(v_{d}-v_{{\rm gas}})-\alpha\eta v_{d}\label{eq:bal_force}
\end{equation}

Solving for the droplet velocity yields: 

\begin{equation}
v_{d}=v_{{\rm gas}}\gamma_{{\rm gas}}/(\gamma_{{\rm gas}}+\eta\alpha)+n_{{\rm drop}}f_{{\rm ext}}/(\gamma_{{\rm gas}}+\eta\alpha)\label{eq:friction}
\end{equation}

By plotting the velocity of the droplet ($v_{d}$) versus the surrounding
gas velocity ($v_{g}$), we validate the drag model (see Inset of
Figure \ref{dropVel_vs_L_kb0}). $v_{d}$ varies linearly with $v_{g}$,
as given by Eq. \ref{eq:friction}. Following Eq. \ref{eq:friction},
the friction coefficients can be estimated. We note that the influence
of the gas on the droplet velocity depends on the relative coexistence
densities of the vapor ($\rho_{v}$) and liquid ($\rho_{l}$) phases
at the given temperature. In this case the temperature was fixed at
$T=0.8\varepsilon/k_{B}$, and the coexisting densities ratio $\rho_{l}/\rho_{v}=0.7\sigma^{-3}/0.03\sigma^{-3}\simeq23$.
We remark that the gas-liquid friction can have an important influence
in two phase flows, and should be taken into account in rheological
analysis in nano-channels.

\section{{\large{}Concluding Remarks\label{sec:Concluding-Remarks}}}

We studied a liquid droplet coexisting with its vapor, flowing through
a nano-channel, whose walls are coated with hydrophobic semiflexible
polymer brushes. It is important to note that the droplet does not
wet the brush, being in a Cassie\textendash Baxter state. 

We studied the influence of the bending rigidity of the polymers on
the flow, maintaining the droplet shape and mass constant. We explored
a wide range of bending stiffness, which expressed as the ratio of
persistence $l_{p}$ over contour lengths $l_{c}$ of the polymer
chains, belong to the range: $0.1<l_{p}/l_{c}<20$. We found that
the center of mass velocity of the droplet decreases with increasing
bending constant. In particular, for the range $l_{p}/l_{c}<3$ the
velocity of the droplet decays faster with increasing bending rigidity,
than for stiffer polymers in the range $l_{p}/l_{c}>3$. In order
to have a better understanding of these results, we investigated the
velocity field of the free terminal bead of the polymers. For flexible
polymers ($l_{p}/l_{c}<3$) this velocity field shows that the monomers
near the liquid move in the same direction as the droplet, while the
terminal beads that are nearer to the substrate move in a direction
opposite to the flow. This collective \textquotedbl{}treadmill-belt\textquotedbl{}
dynamics of the chains gives rise to a positive velocity field of
the brush in contact with the liquid, that allows for a faster droplet
transport through the channel. For $l_{p}/l_{c}>3$ the space explored
by terminal beads of the grafted chains is reduced, due to the bending
rigidity and no cyclic motion is observed, thus reducing the droplet's
velocity. The liquid boundary layer velocity relative to the brush
boundary layer shows a significant increment around $l_{p}/l_{c}\simeq4$.
We attribute this to the decrease in the number of liquid-polymer
interactions, but further investigations should be carried out to
unravel this completely.

We analyzed in addition, the deformation of the polymer brush as a
function of rigidity. The elongation of the chains in the flow direction
presents a maximum for polymers whose persistence length is equal
to the contour length ($l_{p}/l_{c}\simeq1$). For this value of bending
rigidity, the maximum compression relative to the unperturbed brush
is observed. The height of the polymers as a function of persistence
length was also examined, showing a good agreement with mean field
theories \cite{BIRSHTEIN19832165,C4SM02862G} for $l_{p}/l_{c}\leq4$.

For the droplet flowing in coexistence with its gas, we observe that
the velocity of the gas phase decays exponentially as the distance
from the droplet increases. The length of the simulations box was
changed to modify the distance between consecutive droplets, therefore
varying the velocity at which the droplet encounters the gas. The
droplet-vapor friction force was found to depend linearly on the velocity
difference between droplet and its surounding gas, as expected in
a proposed drag model. Reducing the distance between consecutive droplets
increases the velocity of the flow.

Increasing the understanding of droplet flow through brush-covered
surfaces offers a great opportunity to control droplet flow by taking
advantage of the enormous versatility of polymer brushes as surface-modifiers.
This is of great importance for microfluidics and other technological
applications. Hopefully the simulation results presented in this work
can be compared with upcoming experiments. 
\begin{acknowledgement}
Financial support through grants PIP 11220150100417 and PIP 11220110100646
(CONICET), BAPIN 2014, BAPIN 2017 (CNEA), PME 2015, PICT-E 134-2014
(MINCYT) is gratefully acknowledged. We thank also Marcus Müller and
Ignacio Urrutia for fruitful discussions about different aspects of
the present work.
\end{acknowledgement}

\onecolumn

\bibliographystyle{achemso}


\begin{mcitethebibliography}{76}
\providecommand*\natexlab[1]{#1}
\providecommand*\mciteSetBstSublistMode[1]{}
\providecommand*\mciteSetBstMaxWidthForm[2]{}
\providecommand*\mciteBstWouldAddEndPuncttrue
  {\def\EndOfBibitem{\unskip.}}
\providecommand*\mciteBstWouldAddEndPunctfalse
  {\let\EndOfBibitem\relax}
\providecommand*\mciteSetBstMidEndSepPunct[3]{}
\providecommand*\mciteSetBstSublistLabelBeginEnd[3]{}
\providecommand*\EndOfBibitem{}
\mciteSetBstSublistMode{f}
\mciteSetBstMaxWidthForm{subitem}{(\alph{mcitesubitemcount})}
\mciteSetBstSublistLabelBeginEnd
  {\mcitemaxwidthsubitemform\space}
  {\relax}
  {\relax}

\bibitem[Weinbaum \latin{et~al.}(2007)Weinbaum, Tarbell, and
  Damiano]{doi:10.1146/annurev.bioeng.9.060906.151959}
Weinbaum,~S.; Tarbell,~J.~M.; Damiano,~E.~R. The Structure and Function of the
  Endothelial Glycocalyx Layer. \emph{Annual Review of Biomedical Engineering}
  \textbf{2007}, \emph{9}, 121--167\relax
\mciteBstWouldAddEndPuncttrue
\mciteSetBstMidEndSepPunct{\mcitedefaultmidpunct}
{\mcitedefaultendpunct}{\mcitedefaultseppunct}\relax
\EndOfBibitem
\bibitem[Lanotte \latin{et~al.}(2014)Lanotte, Tomaiuolo, Misbah, Bureau, and
  Guido]{Lanotte_2014}
Lanotte,~L.; Tomaiuolo,~G.; Misbah,~C.; Bureau,~L.; Guido,~S. Red blood cell
  dynamics in polymer brush-coated microcapillaries: A model of endothelial
  glycocalyx in vitro. \emph{Biomicrofluidics} \textbf{2014}, \emph{8},
  014104\relax
\mciteBstWouldAddEndPuncttrue
\mciteSetBstMidEndSepPunct{\mcitedefaultmidpunct}
{\mcitedefaultendpunct}{\mcitedefaultseppunct}\relax
\EndOfBibitem
\bibitem[Cruz-Chu \latin{et~al.}(2014)Cruz-Chu, Malafeev, Pajarskas, Pivkin,
  and Koumoutsakos]{CruzChu2014232}
Cruz-Chu,~E.~R.; Malafeev,~A.; Pajarskas,~T.; Pivkin,~I.~V.; Koumoutsakos,~P.
  Structure and Response to Flow of the Glycocalyx Layer. \emph{Biophysical
  Journal} \textbf{2014}, \emph{106}, 232 -- 243\relax
\mciteBstWouldAddEndPuncttrue
\mciteSetBstMidEndSepPunct{\mcitedefaultmidpunct}
{\mcitedefaultendpunct}{\mcitedefaultseppunct}\relax
\EndOfBibitem
\bibitem[Sackmann \latin{et~al.}(2014)Sackmann, Fulton, and
  Beebe]{http://dx.doi.org/10.1038/nature13118}
Sackmann,~E.~K.; Fulton,~A.~L.; Beebe,~D.~J. The present and future role of
  microfluidics in biomedical research. \emph{Nature} \textbf{2014},
  \emph{507}, 181 -- 189\relax
\mciteBstWouldAddEndPuncttrue
\mciteSetBstMidEndSepPunct{\mcitedefaultmidpunct}
{\mcitedefaultendpunct}{\mcitedefaultseppunct}\relax
\EndOfBibitem
\bibitem[Juang and Chang(2016)Juang, and Chang]{BIOT:BIOT201500278}
Juang,~Y.-J.; Chang,~J.-S. Applications of microfluidics in microalgae
  biotechnology: A review. \emph{Biotechnology Journal} \textbf{2016},
  \emph{11}, 327--335\relax
\mciteBstWouldAddEndPuncttrue
\mciteSetBstMidEndSepPunct{\mcitedefaultmidpunct}
{\mcitedefaultendpunct}{\mcitedefaultseppunct}\relax
\EndOfBibitem
\bibitem[Volpatti and Yetisen(2014)Volpatti, and Yetisen]{Volpatti2014347}
Volpatti,~L.~R.; Yetisen,~A.~K. Commercialization of microfluidic devices.
  \emph{Trends in Biotechnology} \textbf{2014}, \emph{32}, 347 -- 350\relax
\mciteBstWouldAddEndPuncttrue
\mciteSetBstMidEndSepPunct{\mcitedefaultmidpunct}
{\mcitedefaultendpunct}{\mcitedefaultseppunct}\relax
\EndOfBibitem
\bibitem[Song(2006)]{12ref19H}
Song,~H. Reactions in Droplets in Microfluidic Channels. \emph{Angewandte
  Chemie (International ed. in English)} \textbf{2006}, \emph{45},
  7336--7356\relax
\mciteBstWouldAddEndPuncttrue
\mciteSetBstMidEndSepPunct{\mcitedefaultmidpunct}
{\mcitedefaultendpunct}{\mcitedefaultseppunct}\relax
\EndOfBibitem
\bibitem[Lauga \latin{et~al.}(2007)Lauga, Brenner, and Stone]{Lauga2007}
Lauga,~E.; Brenner,~M.; Stone,~H. \emph{Springer Handbook of Experimental Fluid
  Mechanics}; Springer Berlin Heidelberg, 2007; pp 1219--1240\relax
\mciteBstWouldAddEndPuncttrue
\mciteSetBstMidEndSepPunct{\mcitedefaultmidpunct}
{\mcitedefaultendpunct}{\mcitedefaultseppunct}\relax
\EndOfBibitem
\bibitem[Yen and Soong(2016)Yen, and Soong]{doi:10.1080/00268976.2015.1119899}
Yen,~T.-H.; Soong,~C.-Y. Effective boundary slip and wetting characteristics of
  water on substrates with effects of surface morphology. \emph{Molecular
  Physics} \textbf{2016}, \emph{114}, 797--809\relax
\mciteBstWouldAddEndPuncttrue
\mciteSetBstMidEndSepPunct{\mcitedefaultmidpunct}
{\mcitedefaultendpunct}{\mcitedefaultseppunct}\relax
\EndOfBibitem
\bibitem[Cao \latin{et~al.}(2006)Cao, Chen, and Guo]{PhysRevE.74.066311}
Cao,~B.-Y.; Chen,~M.; Guo,~Z.-Y. Liquid flow in surface-nanostructured channels
  studied by molecular dynamics simulation. \emph{Phys. Rev. E} \textbf{2006},
  \emph{74}, 066311\relax
\mciteBstWouldAddEndPuncttrue
\mciteSetBstMidEndSepPunct{\mcitedefaultmidpunct}
{\mcitedefaultendpunct}{\mcitedefaultseppunct}\relax
\EndOfBibitem
\bibitem[Zhang(2016)]{Zhang2016295}
Zhang,~Y. Effect of wall surface roughness on mass transfer in a nano channel.
  \emph{International Journal of Heat and Mass Transfer} \textbf{2016},
  \emph{100}, 295 -- 302\relax
\mciteBstWouldAddEndPuncttrue
\mciteSetBstMidEndSepPunct{\mcitedefaultmidpunct}
{\mcitedefaultendpunct}{\mcitedefaultseppunct}\relax
\EndOfBibitem
\bibitem[Yen(2014)]{Yen2014}
Yen,~T.-H. Molecular dynamics simulation of fluid containing gas in hydrophilic
  rough wall nanochannels. \emph{Microfluidics and Nanofluidics} \textbf{2014},
  \emph{17}, 325--339\relax
\mciteBstWouldAddEndPuncttrue
\mciteSetBstMidEndSepPunct{\mcitedefaultmidpunct}
{\mcitedefaultendpunct}{\mcitedefaultseppunct}\relax
\EndOfBibitem
\bibitem[Choi and Kim(2006)Choi, and Kim]{Choi_06}
Choi,~C.-H.; Kim,~C.-J. Large Slip of Aqueous Liquid Flow over a Nanoengineered
  Superhydrophobic Surface. \emph{Phys. Rev. Lett.} \textbf{2006}, \emph{96},
  066001\relax
\mciteBstWouldAddEndPuncttrue
\mciteSetBstMidEndSepPunct{\mcitedefaultmidpunct}
{\mcitedefaultendpunct}{\mcitedefaultseppunct}\relax
\EndOfBibitem
\bibitem[Bixler and Bhushan(2012)Bixler, and Bhushan]{Bixler_12}
Bixler,~G.~D.; Bhushan,~B. Bioinspired rice leaf and butterfly wing surface
  structures combining shark skin and lotus effects. \emph{Soft Matter}
  \textbf{2012}, \emph{8}, 11271--11284\relax
\mciteBstWouldAddEndPuncttrue
\mciteSetBstMidEndSepPunct{\mcitedefaultmidpunct}
{\mcitedefaultendpunct}{\mcitedefaultseppunct}\relax
\EndOfBibitem
\bibitem[Zhang \latin{et~al.}(2012)Zhang, Xia, Kim, and Sun]{Zhang_12}
Zhang,~Y.-L.; Xia,~H.; Kim,~E.; Sun,~H.-B. Recent developments in
  superhydrophobic surfaces with unique structural and functional properties.
  \emph{Soft Matter} \textbf{2012}, \emph{8}, 11217--11231\relax
\mciteBstWouldAddEndPuncttrue
\mciteSetBstMidEndSepPunct{\mcitedefaultmidpunct}
{\mcitedefaultendpunct}{\mcitedefaultseppunct}\relax
\EndOfBibitem
\bibitem[Kunert and Harting(2007)Kunert, and Harting]{PhysRevLett.99.176001}
Kunert,~C.; Harting,~J. Roughness Induced Boundary Slip in Microchannel Flows.
  \emph{Phys. Rev. Lett.} \textbf{2007}, \emph{99}, 176001\relax
\mciteBstWouldAddEndPuncttrue
\mciteSetBstMidEndSepPunct{\mcitedefaultmidpunct}
{\mcitedefaultendpunct}{\mcitedefaultseppunct}\relax
\EndOfBibitem
\bibitem[Liakopoulos \latin{et~al.}(2016)Liakopoulos, Sofos, and
  Karakasidis]{Liakopoulos2016}
Liakopoulos,~A.; Sofos,~F.; Karakasidis,~T.~E. Friction factor in nanochannel
  flows. \emph{Microfluidics and Nanofluidics} \textbf{2016}, \emph{20},
  1--7\relax
\mciteBstWouldAddEndPuncttrue
\mciteSetBstMidEndSepPunct{\mcitedefaultmidpunct}
{\mcitedefaultendpunct}{\mcitedefaultseppunct}\relax
\EndOfBibitem
\bibitem[Tretheway and Meinhart(2004)Tretheway, and
  Meinhart]{:/content/aip/journal/pof2/16/5/10.1063/1.1669400}
Tretheway,~D.~C.; Meinhart,~C.~D. A generating mechanism for apparent fluid
  slip in hydrophobic microchannels. \emph{Physics of Fluids} \textbf{2004},
  \emph{16}, 1509--1515\relax
\mciteBstWouldAddEndPuncttrue
\mciteSetBstMidEndSepPunct{\mcitedefaultmidpunct}
{\mcitedefaultendpunct}{\mcitedefaultseppunct}\relax
\EndOfBibitem
\bibitem[Lichter \latin{et~al.}(2007)Lichter, Martini, Snurr, and
  Wang]{PhysRevLett.98.226001}
Lichter,~S.; Martini,~A.; Snurr,~R.~Q.; Wang,~Q. Liquid Slip in Nanoscale
  Channels as a Rate Process. \emph{Phys. Rev. Lett.} \textbf{2007}, \emph{98},
  226001\relax
\mciteBstWouldAddEndPuncttrue
\mciteSetBstMidEndSepPunct{\mcitedefaultmidpunct}
{\mcitedefaultendpunct}{\mcitedefaultseppunct}\relax
\EndOfBibitem
\bibitem[Fukushima \latin{et~al.}(2015)Fukushima, Mima, Kinefuchi, and
  Tokumasu]{doi:10.1021/acs.jpcc.5b07951}
Fukushima,~A.; Mima,~T.; Kinefuchi,~I.; Tokumasu,~T. Molecular Dynamics
  Simulation of Channel Size Dependence of the Friction Coefficient between a
  Water Droplet and a Nanochannel Wall. \emph{The Journal of Physical Chemistry
  C} \textbf{2015}, \emph{119}, 28396--28404\relax
\mciteBstWouldAddEndPuncttrue
\mciteSetBstMidEndSepPunct{\mcitedefaultmidpunct}
{\mcitedefaultendpunct}{\mcitedefaultseppunct}\relax
\EndOfBibitem
\bibitem[He \latin{et~al.}(2010)He, Hasegawa, and Kasagi]{HE2010126}
He,~Q.; Hasegawa,~Y.; Kasagi,~N. Heat transfer modelling of gas liquid slug
  flow without phase changein a micro tube. \emph{International Journal of Heat
  and Fluid Flow} \textbf{2010}, \emph{31}, 126 -- 136\relax
\mciteBstWouldAddEndPuncttrue
\mciteSetBstMidEndSepPunct{\mcitedefaultmidpunct}
{\mcitedefaultendpunct}{\mcitedefaultseppunct}\relax
\EndOfBibitem
\bibitem[Wu \latin{et~al.}(2013)Wu, Xu, and Qian]{0953-8984-25-19-195103}
Wu,~C.; Xu,~X.; Qian,~T. Molecular dynamics simulations for the motion of
  evaporative droplets driven by thermal gradients along nanochannels.
  \emph{Journal of Physics: Condensed Matter} \textbf{2013}, \emph{25},
  195103\relax
\mciteBstWouldAddEndPuncttrue
\mciteSetBstMidEndSepPunct{\mcitedefaultmidpunct}
{\mcitedefaultendpunct}{\mcitedefaultseppunct}\relax
\EndOfBibitem
\bibitem[G{\"u}nther \latin{et~al.}(2005)G{\"u}nther, Jhunjhunwala, Thalmann,
  Schmidt, and Jensen]{doi:10.1021/la0482406}
G{\"u}nther,~A.; Jhunjhunwala,~M.; Thalmann,~M.; Schmidt,~M.~A.; Jensen,~K.~F.
  Micromixing of Miscible Liquids in Segmented Gas Liquid Flow. \emph{Langmuir}
  \textbf{2005}, \emph{21}, 1547--1555\relax
\mciteBstWouldAddEndPuncttrue
\mciteSetBstMidEndSepPunct{\mcitedefaultmidpunct}
{\mcitedefaultendpunct}{\mcitedefaultseppunct}\relax
\EndOfBibitem
\bibitem[Chen \latin{et~al.}(2014)Chen, Li, Zhou, Pelan, Stoyanov, Arnaudov,
  and Stone]{doi:10.1021/la5004929}
Chen,~H.; Li,~J.; Zhou,~W.; Pelan,~E.~G.; Stoyanov,~S.~D.; Arnaudov,~L.~N.;
  Stone,~H.~A. Sonication Microfluidics for Fabrication of
  Nanoparticle-Stabilized Microbubbles. \emph{Langmuir} \textbf{2014},
  \emph{30}, 4262--4266 \relax
\mciteBstWouldAddEndPuncttrue
\mciteSetBstMidEndSepPunct{\mcitedefaultmidpunct}
{\mcitedefaultendpunct}{\mcitedefaultseppunct}\relax
\EndOfBibitem
\bibitem[Kelly \latin{et~al.}(2015)Kelly, Balhoff, and
  Torres-Verd{\'i}n]{doi:10.1021/la504742w}
Kelly,~S.; Balhoff,~M.~T.; Torres-Verd{\'i}n,~C. Quantification of Bulk
  Solution Limits for Liquid and Interfacial Transport in Nanoconfinements.
  \emph{Langmuir} \textbf{2015}, \emph{31}, 2167--2179 \relax
\mciteBstWouldAddEndPuncttrue
\mciteSetBstMidEndSepPunct{\mcitedefaultmidpunct}
{\mcitedefaultendpunct}{\mcitedefaultseppunct}\relax
\EndOfBibitem
\bibitem[Kuhn \latin{et~al.}(2011)Kuhn, Hartman, Sultana, Nagy, Marre, and
  Jensen]{doi:10.1021/la2004744}
Kuhn,~S.; Hartman,~R.~L.; Sultana,~M.; Nagy,~K.~D.; Marre,~S.; Jensen,~K.~F.
  Teflon-Coated Silicon Microreactors: Impact on Segmented Liquid-Liquid
  Multiphase Flows. \emph{Langmuir} \textbf{2011}, \emph{27}, 6519--6527 \relax
\mciteBstWouldAddEndPuncttrue
\mciteSetBstMidEndSepPunct{\mcitedefaultmidpunct}
{\mcitedefaultendpunct}{\mcitedefaultseppunct}\relax
\EndOfBibitem
\bibitem[Mawatari \latin{et~al.}(2012)Mawatari, Kubota, Xu, Priest, Sedev,
  Ralston, and Kitamori]{doi:10.1021/ac3028905}
Mawatari,~K.; Kubota,~S.; Xu,~Y.; Priest,~C.; Sedev,~R.; Ralston,~J.;
  Kitamori,~T. Femtoliter Droplet Handling in Nanofluidic Channels: A Laplace
  Nanovalve. \emph{Analytical Chemistry} \textbf{2012}, \emph{84},
  10812--10816 \relax
\mciteBstWouldAddEndPuncttrue
\mciteSetBstMidEndSepPunct{\mcitedefaultmidpunct}
{\mcitedefaultendpunct}{\mcitedefaultseppunct}\relax
\EndOfBibitem
\bibitem[Shui \latin{et~al.}(2007)Shui, Eijkel, and van~den Berg]{Shui200735}
Shui,~L.; Eijkel,~J.~C.; van~den Berg,~A. Multiphase flow in microfluidic
  systems Control and applications of droplets and interfaces. \emph{Advances
  in Colloid and Interface Science} \textbf{2007}, \emph{133}, 35 -- 49\relax
\mciteBstWouldAddEndPuncttrue
\mciteSetBstMidEndSepPunct{\mcitedefaultmidpunct}
{\mcitedefaultendpunct}{\mcitedefaultseppunct}\relax
\EndOfBibitem
\bibitem[Rebrov(2010)]{Rebrov2010}
Rebrov,~E.~V. Two-phase flow regimes in microchannels. \emph{Theoretical
  Foundations of Chemical Engineering} \textbf{2010}, \emph{44}, 355--367\relax
\mciteBstWouldAddEndPuncttrue
\mciteSetBstMidEndSepPunct{\mcitedefaultmidpunct}
{\mcitedefaultendpunct}{\mcitedefaultseppunct}\relax
\EndOfBibitem
\bibitem[Zhang \latin{et~al.}(2006)Zhang, Yang, and Wang]{Zhang01022006}
Zhang,~F.~Y.; Yang,~X.~G.; Wang,~C.~Y. Liquid Water Removal from a Polymer
  Electrolyte Fuel Cell. \emph{Journal of The Electrochemical Society}
  \textbf{2006}, \emph{153}, A225--A232\relax
\mciteBstWouldAddEndPuncttrue
\mciteSetBstMidEndSepPunct{\mcitedefaultmidpunct}
{\mcitedefaultendpunct}{\mcitedefaultseppunct}\relax
\EndOfBibitem
\bibitem[Lu \latin{et~al.}(2009)Lu, Kandlikar, Rath, Grimm, Domigan, White,
  Hardbarger, Owejan, and Trabold]{Lu20093445}
Lu,~Z.; Kandlikar,~S.; Rath,~C.; Grimm,~M.; Domigan,~W.; White,~A.;
  Hardbarger,~M.; Owejan,~J.; Trabold,~T. Water management studies in PEM
  fuel cells, Part II: Ex situ investigation of flow maldistribution, pressure
  drop and two-phase flow pattern in gas channels. \emph{International Journal
  of Hydrogen Energy} \textbf{2009}, \emph{34}, 3445 -- 3456\relax
\mciteBstWouldAddEndPuncttrue
\mciteSetBstMidEndSepPunct{\mcitedefaultmidpunct}
{\mcitedefaultendpunct}{\mcitedefaultseppunct}\relax
\EndOfBibitem
\bibitem[Cheah \latin{et~al.}(2013)Cheah, Kevrekidis, and
  Benziger]{doi:10.1021/la403057k}
Cheah,~M.~J.; Kevrekidis,~I.~G.; Benziger,~J.~B. Water Slug to Drop and Film
  Transitions in Gas-Flow Channels. \emph{Langmuir} \textbf{2013}, \emph{29},
  15122--15136 \relax
\mciteBstWouldAddEndPuncttrue
\mciteSetBstMidEndSepPunct{\mcitedefaultmidpunct}
{\mcitedefaultendpunct}{\mcitedefaultseppunct}\relax
\EndOfBibitem
\bibitem[Seemann \latin{et~al.}(2012)Seemann, Brinkmann, Pfohl, and
  Herminghaus]{10.1088/0034-4885/75/1/016601}
Seemann,~R.; Brinkmann,~M.; Pfohl,~T.; Herminghaus,~S. Droplet based
  microfluidics. \emph{Rep. Prog. Phys.} \textbf{2012}, \emph{75}, 016601\relax
\mciteBstWouldAddEndPuncttrue
\mciteSetBstMidEndSepPunct{\mcitedefaultmidpunct}
{\mcitedefaultendpunct}{\mcitedefaultseppunct}\relax
\EndOfBibitem
\bibitem[Wei \latin{et~al.}(2013)Wei, Cai, Zhou, and
  Liu]{doi:10.1021/ma401537j}
Wei,~Q.; Cai,~M.; Zhou,~F.; Liu,~W. Dramatically Tuning Friction Using
  Responsive Polyelectrolyte Brushes. \emph{Macromolecules} \textbf{2013},
  \emph{46}, 9368--9379\relax
\mciteBstWouldAddEndPuncttrue
\mciteSetBstMidEndSepPunct{\mcitedefaultmidpunct}
{\mcitedefaultendpunct}{\mcitedefaultseppunct}\relax
\EndOfBibitem
\bibitem[Chen \latin{et~al.}(2010)Chen, Ferris, Zhang, Ducker, and
  Zauscher]{Chen201094}
Chen,~T.; Ferris,~R.; Zhang,~J.; Ducker,~R.; Zauscher,~S. Stimulus-responsive
  polymer brushes on surfaces: Transduction mechanisms and applications.
  \emph{Progress in Polymer Science} \textbf{2010}, \emph{35}, 94 -- 112 \relax
\mciteBstWouldAddEndPuncttrue
\mciteSetBstMidEndSepPunct{\mcitedefaultmidpunct}
{\mcitedefaultendpunct}{\mcitedefaultseppunct}\relax
\EndOfBibitem
\bibitem[Chen \latin{et~al.}(2017)Chen, Cordero, Tran, and
  Ober]{doi:10.1021/acs.macromol.7b00450}
Chen,~W.-L.; Cordero,~R.; Tran,~H.; Ober,~C.~K. 50th Anniversary Perspective:
  Polymer Brushes: Novel Surfaces for Future Materials. \emph{Macromolecules}
  \textbf{2017}, \emph{50}, 4089--4113\relax
\mciteBstWouldAddEndPuncttrue
\mciteSetBstMidEndSepPunct{\mcitedefaultmidpunct}
{\mcitedefaultendpunct}{\mcitedefaultseppunct}\relax
\EndOfBibitem
\bibitem[Ma \latin{et~al.}(2015)Ma, Wang, Liang, Sun, Gorb, and
  Zhou]{SMLL:SMLL201402484}
Ma,~S.; Wang,~D.; Liang,~Y.; Sun,~B.; Gorb,~S.~N.; Zhou,~F. Gecko-Inspired but
  Chemically Switched Friction and Adhesion on Nanofibrillar Surfaces.
  \emph{Small} \textbf{2015}, \emph{11}, 1131--1137\relax
\mciteBstWouldAddEndPuncttrue
\mciteSetBstMidEndSepPunct{\mcitedefaultmidpunct}
{\mcitedefaultendpunct}{\mcitedefaultseppunct}\relax
\EndOfBibitem
\bibitem[Azzaroni(2012)]{POLA:POLA26119}
Azzaroni,~O. Polymer brushes here, there, and everywhere: Recent advances in
  their practical applications and emerging opportunities in multiple research
  fields. \emph{Journal of Polymer Science Part A: Polymer Chemistry}
  \textbf{2012}, \emph{50}, 3225--3258\relax
\mciteBstWouldAddEndPuncttrue
\mciteSetBstMidEndSepPunct{\mcitedefaultmidpunct}
{\mcitedefaultendpunct}{\mcitedefaultseppunct}\relax
\EndOfBibitem
\bibitem[Das \latin{et~al.}(2015)Das, Banik, Chen, Sinha, and
  Mukherjee]{C5SM01962A}
Das,~S.; Banik,~M.; Chen,~G.; Sinha,~S.; Mukherjee,~R. Polyelectrolyte brushes:
  theory{,} modelling{,} synthesis and applications. \emph{Soft Matter}
  \textbf{2015}, \emph{11}, 8550--8583\relax
\mciteBstWouldAddEndPuncttrue
\mciteSetBstMidEndSepPunct{\mcitedefaultmidpunct}
{\mcitedefaultendpunct}{\mcitedefaultseppunct}\relax
\EndOfBibitem
\bibitem[Yameen \latin{et~al.}(2009)Yameen, Ali, Neumann, Ensinger, Knoll, and
  Azzaroni]{doi:10.1021/ja8086104}
Yameen,~B.; Ali,~M.; Neumann,~R.; Ensinger,~W.; Knoll,~W.; Azzaroni,~O. Single
  Conical Nanopores Displaying pH-Tunable Rectifying Characteristics.
  Manipulating Ionic Transport With Zwitterionic Polymer Brushes. \emph{Journal
  of the American Chemical Society} \textbf{2009}, \emph{131}, 2070--2071,
  \relax
\mciteBstWouldAddEndPuncttrue
\mciteSetBstMidEndSepPunct{\mcitedefaultmidpunct}
{\mcitedefaultendpunct}{\mcitedefaultseppunct}\relax
\EndOfBibitem
\bibitem[Costantini \latin{et~al.}(2009)Costantini, Bula, Salvio, Huskens,
  Gardeniers, Reinhoudt, and Verboom]{doi:10.1021/ja807616z}
Costantini,~F.; Bula,~W.~P.; Salvio,~R.; Huskens,~J.; Gardeniers,~H. J. G.~E.;
  Reinhoudt,~D.~N.; Verboom,~W. Nanostructure Based on Polymer Brushes for
  Efficient Heterogeneous Catalysis in Microreactors. \emph{Journal of the
  American Chemical Society} \textbf{2009}, \emph{131}, 1650--1651\relax
\mciteBstWouldAddEndPuncttrue
\mciteSetBstMidEndSepPunct{\mcitedefaultmidpunct}
{\mcitedefaultendpunct}{\mcitedefaultseppunct}\relax
\EndOfBibitem
\bibitem[Yu \latin{et~al.}(2010)Yu, Zhang, Chen, Zhou, Wu, Huang, and
  Brash]{doi:10.1021/la904663m}
Yu,~Q.; Zhang,~Y.; Chen,~H.; Zhou,~F.; Wu,~Z.; Huang,~H.; Brash,~J.~L. Protein
  Adsorption and Cell Adhesion/Detachment Behavior on Dual-Responsive Silicon
  Surfaces Modified with Poly(N-isopropylacrylamide)-block-polystyrene
  Copolymer. \emph{Langmuir} \textbf{2010}, \emph{26}, 8582--8588 
  \relax
\mciteBstWouldAddEndPuncttrue
\mciteSetBstMidEndSepPunct{\mcitedefaultmidpunct}
{\mcitedefaultendpunct}{\mcitedefaultseppunct}\relax
\EndOfBibitem
\bibitem[Pastorino \latin{et~al.}(2009)Pastorino, Binder, and
  M{\"u}ller]{doi:10.1021/ma8015757}
Pastorino,~C.; Binder,~K.; M{\"u}ller,~M. Coarse-Grained Description of a
  Brush-Melt Interface in Equilibrium and under Flow. \emph{Macromolecules}
  \textbf{2009}, \emph{42}, 401--410\relax
\mciteBstWouldAddEndPuncttrue
\mciteSetBstMidEndSepPunct{\mcitedefaultmidpunct}
{\mcitedefaultendpunct}{\mcitedefaultseppunct}\relax
\EndOfBibitem
\bibitem[Lee \latin{et~al.}(2012)Lee, Hendy, and Neto]{Lee_12}
Lee,~T.; Hendy,~S.~C.; Neto,~C. Interfacial Flow of Simple Liquids on Polymer
  Brushes: Effect of Solvent Quality and Grafting Density.
  \emph{Macromolecules} \textbf{2012}, \emph{45}, 6241--6252\relax
\mciteBstWouldAddEndPuncttrue
\mciteSetBstMidEndSepPunct{\mcitedefaultmidpunct}
{\mcitedefaultendpunct}{\mcitedefaultseppunct}\relax
\EndOfBibitem
\bibitem[Sokoloff(2007)]{doi:10.1021/ma062875p}
Sokoloff,~J.~B. Theory of Friction between Neutral Polymer Brushes.
  \emph{Macromolecules} \textbf{2007}, \emph{40}, 4053--4058\relax
\mciteBstWouldAddEndPuncttrue
\mciteSetBstMidEndSepPunct{\mcitedefaultmidpunct}
{\mcitedefaultendpunct}{\mcitedefaultseppunct}\relax
\EndOfBibitem
\bibitem[Desai \latin{et~al.}(2017)Desai, Sinha, and Das]{C7SM00466D}
Desai,~P.~R.; Sinha,~S.; Das,~S. Compression of polymer brushes in the weak
  interpenetration regime: scaling theory and molecular dynamics simulations.
  \emph{Soft Matter} \textbf{2017}, 4159--4166\relax
\mciteBstWouldAddEndPuncttrue
\mciteSetBstMidEndSepPunct{\mcitedefaultmidpunct}
{\mcitedefaultendpunct}{\mcitedefaultseppunct}\relax
\EndOfBibitem
\bibitem[Zhu \latin{et~al.}(2002)Zhu, , and Granick]{doi:10.1021/ma020043v}
Zhu,~Y.; ; Granick,~S. Apparent Slip of Newtonian Fluids Past Adsorbed Polymer
  Layers. \emph{Macromolecules} \textbf{2002}, \emph{35}, 4658--4663\relax
\mciteBstWouldAddEndPuncttrue
\mciteSetBstMidEndSepPunct{\mcitedefaultmidpunct}
{\mcitedefaultendpunct}{\mcitedefaultseppunct}\relax
\EndOfBibitem
\bibitem[Klein \latin{et~al.}(1994)Klein, Kumacheva, Mahalu, Perahia, and
  Fetters]{10.1038/370634a0}
Klein,~J.; Kumacheva,~E.; Mahalu,~D.; Perahia,~D.; Fetters,~L.~J. Reduction of
  frictional forces between solid surfaces bearing polymer brushes.
  \emph{Nature} \textbf{1994}, \emph{370}, 634--636\relax
\mciteBstWouldAddEndPuncttrue
\mciteSetBstMidEndSepPunct{\mcitedefaultmidpunct}
{\mcitedefaultendpunct}{\mcitedefaultseppunct}\relax
\EndOfBibitem
\bibitem[Zhang \latin{et~al.}(2015)Zhang, Ma, Wei, Ye, Yu, van~der Gucht, and
  Zhou]{doi:10.1021/acs.macromol.5b01267}
Zhang,~R.; Ma,~S.; Wei,~Q.; Ye,~Q.; Yu,~B.; van~der Gucht,~J.; Zhou,~F. The
  Weak Interaction of Surfactants with Polymer Brushes and Its Impact on
  Lubricating Behavior. \emph{Macromolecules} \textbf{2015}, \emph{48},
  6186--6196\relax
\mciteBstWouldAddEndPuncttrue
\mciteSetBstMidEndSepPunct{\mcitedefaultmidpunct}
{\mcitedefaultendpunct}{\mcitedefaultseppunct}\relax
\EndOfBibitem
\bibitem[Charrault \latin{et~al.}(2016)Charrault, Lee, Easton, and
  Neto]{C5SM02546J}
Charrault,~E.; Lee,~T.; Easton,~C.~D.; Neto,~C. Boundary flow on end-grafted
  PEG brushes. \emph{Soft Matter} \textbf{2016}, \emph{12}, 1906--1914\relax
\mciteBstWouldAddEndPuncttrue
\mciteSetBstMidEndSepPunct{\mcitedefaultmidpunct}
{\mcitedefaultendpunct}{\mcitedefaultseppunct}\relax
\EndOfBibitem
\bibitem[Chennevi{\'e}re \latin{et~al.}(2016)Chennevi{\'e}re, Cousin, Bou{\'e},
  Drockenmuller, Shull, L{\'e}ger, and
  Restagno]{doi:10.1021/acs.macromol.5b02505}
Chennevi{\'e}re,~A.; Cousin,~F.; Bou{\'e},~F.; Drockenmuller,~E.; Shull,~K.~R.;
  L{\'e}ger,~L.; Restagno,~F. Direct Molecular Evidence of the Origin of Slip
  of Polymer Melts on Grafted Brushes. \emph{Macromolecules} \textbf{2016},
  \emph{49}, 2348--2353\relax
\mciteBstWouldAddEndPuncttrue
\mciteSetBstMidEndSepPunct{\mcitedefaultmidpunct}
{\mcitedefaultendpunct}{\mcitedefaultseppunct}\relax
\EndOfBibitem
\bibitem[Deng \latin{et~al.}(2012)Deng, Li, Liang, Caswell, and
  Karniadakis]{Deng_2012}
Deng,~M.; Li,~X.; Liang,~H.; Caswell,~B.; Karniadakis,~G.~E. Simulation and
  modelling of slip flow over surfaces grafted with polymer brushes and
  glycocalyx fibres. \emph{Journal of Fluid Mechanics} \textbf{2012},
  \emph{711}, 192--211\relax
\mciteBstWouldAddEndPuncttrue
\mciteSetBstMidEndSepPunct{\mcitedefaultmidpunct}
{\mcitedefaultendpunct}{\mcitedefaultseppunct}\relax
\EndOfBibitem
\bibitem[Leonforte \latin{et~al.}(2011)Leonforte, Servantie, Pastorino, and
  M{\"u}ller]{0953-8984-23-18-184105}
Leonforte,~F.; Servantie,~J.; Pastorino,~C.; M{\"u}ller,~M. Molecular transport
  and flow past hard and soft surfaces: computer simulation of model systems.
  \emph{Journal of Physics: Condensed Matter} \textbf{2011}, \emph{23},
  184105\relax
\mciteBstWouldAddEndPuncttrue
\mciteSetBstMidEndSepPunct{\mcitedefaultmidpunct}
{\mcitedefaultendpunct}{\mcitedefaultseppunct}\relax
\EndOfBibitem
\bibitem[Priezjev(2009)]{PhysRevE.80.031608}
Priezjev,~N.~V. Shear rate threshold for the boundary slip in dense polymer
  films. \emph{Phys. Rev. E} \textbf{2009}, \emph{80}, 031608\relax
\mciteBstWouldAddEndPuncttrue
\mciteSetBstMidEndSepPunct{\mcitedefaultmidpunct}
{\mcitedefaultendpunct}{\mcitedefaultseppunct}\relax
\EndOfBibitem
\bibitem[Priezjev and Troian(2004)Priezjev, and Troian]{PhysRevLett.92.018302}
Priezjev,~N.~V.; Troian,~S.~M. Molecular Origin and Dynamic Behavior of Slip in
  Sheared Polymer Films. \emph{Phys. Rev. Lett.} \textbf{2004}, \emph{92},
  018302\relax
\mciteBstWouldAddEndPuncttrue
\mciteSetBstMidEndSepPunct{\mcitedefaultmidpunct}
{\mcitedefaultendpunct}{\mcitedefaultseppunct}\relax
\EndOfBibitem
\bibitem[M{\"u}ller \latin{et~al.}(2008)M{\"u}ller, Pastorino, and
  Servantie]{0953-8984-20-49-494225}
M{\"u}ller,~M.; Pastorino,~C.; Servantie,~J. Flow, slippage and a hydrodynamic
  boundary condition of polymers at surfaces. \emph{Journal of Physics:
  Condensed Matter} \textbf{2008}, \emph{20}, 494225\relax
\mciteBstWouldAddEndPuncttrue
\mciteSetBstMidEndSepPunct{\mcitedefaultmidpunct}
{\mcitedefaultendpunct}{\mcitedefaultseppunct}\relax
\EndOfBibitem
\bibitem[M{\"u}ller and MacDowell(2000)M{\"u}ller, and
  MacDowell]{doi:10.1021/ma991796t}
M{\"u}ller,~M.; MacDowell,~L.~G. Interface and Surface Properties of Short
  Polymers in Solution: Monte Carlo Simulations and Self-Consistent Field
  Theory. \emph{Macromolecules} \textbf{2000}, \emph{33}, 3902--3923\relax
\mciteBstWouldAddEndPuncttrue
\mciteSetBstMidEndSepPunct{\mcitedefaultmidpunct}
{\mcitedefaultendpunct}{\mcitedefaultseppunct}\relax
\EndOfBibitem
\bibitem[Kreer(2016)]{C5SM02919H}
Kreer,~T. Polymer-brush lubrication: a review of recent theoretical advances.
  \emph{Soft Matter} \textbf{2016}, \emph{12}, 3479--3501\relax
\mciteBstWouldAddEndPuncttrue
\mciteSetBstMidEndSepPunct{\mcitedefaultmidpunct}
{\mcitedefaultendpunct}{\mcitedefaultseppunct}\relax
\EndOfBibitem
\bibitem[Kim \latin{et~al.}(2009)Kim, Lobaskin, Gutsche, Kremer, Pincus, and
  Netz]{Kim_09}
Kim,~Y.~W.; Lobaskin,~V.; Gutsche,~C.; Kremer,~F.; Pincus,~P.; Netz,~R.~R.
  Nonlinear Response of Grafted Semiflexible Polymers in Shear Flow.
  \emph{Macromolecules} \textbf{2009}, \emph{42}, 3650--3655\relax
\mciteBstWouldAddEndPuncttrue
\mciteSetBstMidEndSepPunct{\mcitedefaultmidpunct}
{\mcitedefaultendpunct}{\mcitedefaultseppunct}\relax
\EndOfBibitem
\bibitem[R{\"o}mer and Fedosov(2015)R{\"o}mer, and Fedosov]{Roemer_15}
R{\"o}mer,~F.; Fedosov,~D.~A. Dense brushes of stiff polymers or filaments in
  fluid flow. \emph{EPL (Europhysics Letters)} \textbf{2015}, \emph{109},
  68001\relax
\mciteBstWouldAddEndPuncttrue
\mciteSetBstMidEndSepPunct{\mcitedefaultmidpunct}
{\mcitedefaultendpunct}{\mcitedefaultseppunct}\relax
\EndOfBibitem
\bibitem[Singh \latin{et~al.}(2016)Singh, Ilg, Espinosa-Marzal, Spencer, and
  Kr{\"o}ger]{polym8070254}
Singh,~M.~K.; Ilg,~P.; Espinosa-Marzal,~R.~M.; Spencer,~N.~D.; Kr{\"o}ger,~M.
  Influence of Chain Stiffness, Grafting Density and Normal Load on the
  Tribological and Structural Behavior of Polymer Brushes: A
  Nonequilibrium-Molecular-Dynamics Study. \emph{Polymers} \textbf{2016},
  \emph{8}, 254\relax
\mciteBstWouldAddEndPuncttrue
\mciteSetBstMidEndSepPunct{\mcitedefaultmidpunct}
{\mcitedefaultendpunct}{\mcitedefaultseppunct}\relax
\EndOfBibitem
\bibitem[Speyer and Pastorino(2015)Speyer, and Pastorino]{C5SM01075F}
Speyer,~K.; Pastorino,~C. Brushes of semiflexible polymers in equilibrium and
  under flow in a super-hydrophobic regime. \emph{Soft Matter} \textbf{2015},
  \emph{11}, 5473--5484\relax
\mciteBstWouldAddEndPuncttrue
\mciteSetBstMidEndSepPunct{\mcitedefaultmidpunct}
{\mcitedefaultendpunct}{\mcitedefaultseppunct}\relax
\EndOfBibitem
\bibitem[Grest and Kremer(1986)Grest, and Kremer]{PhysRevA.33.3628}
Grest,~G.~S.; Kremer,~K. Molecular dynamics simulation for polymers in the
  presence of a heat bath. \emph{Phys. Rev. A} \textbf{1986}, \emph{33},
  3628--3631\relax
\mciteBstWouldAddEndPuncttrue
\mciteSetBstMidEndSepPunct{\mcitedefaultmidpunct}
{\mcitedefaultendpunct}{\mcitedefaultseppunct}\relax
\EndOfBibitem
\bibitem[Nikoubashman \latin{et~al.}(2016)Nikoubashman, Milchev, and
  Binder]{doi:10.1063/1.4971861}
Nikoubashman,~A.; Milchev,~A.; Binder,~K. Dynamics of single semiflexible
  polymers in dilute solution. \emph{The Journal of Chemical Physics}
  \textbf{2016}, \emph{145}, 234903\relax
\mciteBstWouldAddEndPuncttrue
\mciteSetBstMidEndSepPunct{\mcitedefaultmidpunct}
{\mcitedefaultendpunct}{\mcitedefaultseppunct}\relax
\EndOfBibitem
\bibitem[{Hoogerbrugge} and {Koelman}(1992){Hoogerbrugge}, and
  {Koelman}]{Hoogerbrugge_92}
{Hoogerbrugge},~P.~J.; {Koelman},~J.~M.~V.~A. {Simulating microscopic
  hydrodynamic phenomena with dissipative particle dynamics}. \emph{EPL
  (Europhysics Letters)} \textbf{1992}, \emph{19}, 155\relax
\mciteBstWouldAddEndPuncttrue
\mciteSetBstMidEndSepPunct{\mcitedefaultmidpunct}
{\mcitedefaultendpunct}{\mcitedefaultseppunct}\relax
\EndOfBibitem
\bibitem[Espa\~nol(1995)]{Espanol_95}
Espa\~nol,~P. Hydrodynamics from dissipative particle dynamics. \emph{Phys.
  Rev. E} \textbf{1995}, \emph{52}, 1734--1742\relax
\mciteBstWouldAddEndPuncttrue
\mciteSetBstMidEndSepPunct{\mcitedefaultmidpunct}
{\mcitedefaultendpunct}{\mcitedefaultseppunct}\relax
\EndOfBibitem
\bibitem[Soddemann \latin{et~al.}(2003)Soddemann, D\"unweg, and
  Kremer]{Soddemann_03}
Soddemann,~T.; D\"unweg,~B.; Kremer,~K. Dissipative particle dynamics: A useful
  thermostat for equilibrium and nonequilibrium molecular dynamics simulations.
  \emph{Phys. Rev. E} \textbf{2003}, \emph{68}, 046702\relax
\mciteBstWouldAddEndPuncttrue
\mciteSetBstMidEndSepPunct{\mcitedefaultmidpunct}
{\mcitedefaultendpunct}{\mcitedefaultseppunct}\relax
\EndOfBibitem
\bibitem[Pastorino \latin{et~al.}(2007)Pastorino, Kreer, M\"uller, and
  Binder]{Pastorino_07}
Pastorino,~C.; Kreer,~T.; M\"uller,~M.; Binder,~K. Comparison of dissipative
  particle dynamics and Langevin thermostats for out-of-equilibrium simulations
  of polymeric systems. \emph{Phys. Rev. E} \textbf{2007}, \emph{76},
  026706\relax
\mciteBstWouldAddEndPuncttrue
\mciteSetBstMidEndSepPunct{\mcitedefaultmidpunct}
{\mcitedefaultendpunct}{\mcitedefaultseppunct}\relax
\EndOfBibitem
\bibitem[Pastorino \latin{et~al.}(2006)Pastorino, Binder, Kreer, and
  M\"{u}ller]{pastorino:064902}
Pastorino,~C.; Binder,~K.; Kreer,~T.; M\"{u}ller,~M. Static and dynamic
  properties of the interface between a polymer brush and a melt of identical
  chains. \emph{The Journal of Chemical Physics} \textbf{2006}, \emph{124},
  064902\relax
\mciteBstWouldAddEndPuncttrue
\mciteSetBstMidEndSepPunct{\mcitedefaultmidpunct}
{\mcitedefaultendpunct}{\mcitedefaultseppunct}\relax
\EndOfBibitem
\bibitem[M{\"u}ller and Pastorino(2008)M{\"u}ller, and Pastorino]{Mueller_08b}
M{\"u}ller,~M.; Pastorino,~C. Cyclic motion and inversion of surface flow
  direction in a dense polymer brush under shear. \emph{EPL (Europhysics
  Letters)} \textbf{2008}, \emph{81}, 28002\relax
\mciteBstWouldAddEndPuncttrue
\mciteSetBstMidEndSepPunct{\mcitedefaultmidpunct}
{\mcitedefaultendpunct}{\mcitedefaultseppunct}\relax
\EndOfBibitem
\bibitem[Pastorino and M{\"u}ller(2014)Pastorino, and
  M{\"u}ller]{:/content/aip/journal/jcp/140/1/10.1063/1.4851195}
Pastorino,~C.; M{\"u}ller,~M. Mixed brush of chemically and physically adsorbed
  polymers under shear: Inverse transport of the physisorbed species. \emph{The
  Journal of Chemical Physics} \textbf{2014}, \emph{140}, 014901\relax
\mciteBstWouldAddEndPuncttrue
\mciteSetBstMidEndSepPunct{\mcitedefaultmidpunct}
{\mcitedefaultendpunct}{\mcitedefaultseppunct}\relax
\EndOfBibitem
\bibitem[Ramakrishna \latin{et~al.}(2017)Ramakrishna, Cirelli, Divandari, and
  Benetti]{doi:10.1021/acs.langmuir.7b00217}
Ramakrishna,~S.~N.; Cirelli,~M.; Divandari,~M.; Benetti,~E.~M. Effects of
  Lateral Deformation by Thermoresponsive Polymer Brushes on the Measured
  Friction Forces. \emph{Langmuir} \textbf{2017}, \emph{33}, 4164--4171\relax
\mciteBstWouldAddEndPuncttrue
\mciteSetBstMidEndSepPunct{\mcitedefaultmidpunct}
{\mcitedefaultendpunct}{\mcitedefaultseppunct}\relax
\EndOfBibitem
\bibitem[Birshtein and Zhulina(1983)Birshtein, and Zhulina]{BIRSHTEIN19832165}
Birshtein,~T.; Zhulina,~Y. Conformations of polymer chains grafted to an
  impermeable plane surface. \emph{Polymer Science U.S.S.R.} \textbf{1983},
  \emph{25}, 2165 -- 2174\relax
\mciteBstWouldAddEndPuncttrue
\mciteSetBstMidEndSepPunct{\mcitedefaultmidpunct}
{\mcitedefaultendpunct}{\mcitedefaultseppunct}\relax
\EndOfBibitem
\bibitem[Milchev and Binder(2014)Milchev, and Binder]{Milchev_14b}
Milchev,~A.; Binder,~K. Unconventional ordering behavior of semi-flexible
  polymers in dense brushes under compression. \emph{Soft Matter}
  \textbf{2014}, \emph{10}, 3783--3797\relax
\mciteBstWouldAddEndPuncttrue
\mciteSetBstMidEndSepPunct{\mcitedefaultmidpunct}
{\mcitedefaultendpunct}{\mcitedefaultseppunct}\relax
\EndOfBibitem
\bibitem[Egorov \latin{et~al.}(2015)Egorov, Hsu, Milchev, and
  Binder]{C4SM02862G}
Egorov,~S.~A.; Hsu,~H.-P.; Milchev,~A.; Binder,~K. Semiflexible polymer brushes
  and the brush-mushroom crossover. \emph{Soft Matter} \textbf{2015},
  \emph{11}, 2604--2616\relax
\mciteBstWouldAddEndPuncttrue
\mciteSetBstMidEndSepPunct{\mcitedefaultmidpunct}
{\mcitedefaultendpunct}{\mcitedefaultseppunct}\relax
\EndOfBibitem
\end{mcitethebibliography}

\providecommand{\latin}[1]{#1}
\providecommand*\mcitethebibliography{\thebibliography}
\csname @ifundefined\endcsname{endmcitethebibliography}
  {\let\endmcitethebibliography\endthebibliography}{}



\bigskip

This document is the unedited Author's version of a Submitted Work that was subsequently accepted for
publication in Langmuir, copyright American Chemical Society after peer review. To access the final edited
and published work see 

http://pubs.acs.org/doi/10.1021/acs.langmuir.7b02640 

(DOI: 10.1021/acs.langmuir.7b02640)

\end{document}